\documentclass[traditabstract]{aa} 
\usepackage{graphicx}
\usepackage{txfonts}
\begin{document}
\title{The neutral gas extent of galaxies as derived
from weak intervening Ca\,{\sc ii} absorbers
\thanks{Based on observations collected at the 
European Organisation for Astronomical Research in the
Southern Hemisphere, Chile}}


\author{
        P. Richter\inst{1},
        F. Krause \inst{1},
        C. Fechner \inst{1},
        J.C. Charlton \inst{2},
        \and
        M.T. Murphy \inst{3}
        }

\institute{Institut f\"ur Physik und Astronomie, Universit\"at Potsdam,
           Karl-Liebknecht-Str.\,24/25, 14476 Golm, Germany\\
           \email{prichter@astro.physik.uni-potsdam.de}
           \and
           Department of Astronomy and Astrophysics, Pennsylvania State
           University, University Park, PA 16802, USA
           \and
           Centre for Astrophysics \& Supercomputing, Swinburne
           University of Technology, Hawthorn, Victoria 3122, Australia
           }

\date{Received 11 August 2010 / Accepted 03 December 2010}

\abstract{
We present a systematic study of weak intervening Ca\,{\sc ii} absorbers at low
redshift ($z<0.5$), based on the analysis of archival high-resolution ($R\geq 45,000$)
optical spectra of 304 quasars and active galactic nuclei observed with VLT/UVES.
Along a total redshift path of $\Delta z\approx100$ we detected 23 intervening
Ca\,{\sc ii} absorbers in both the Ca\,{\sc ii} H \& K lines, with
rest frame equivalent widths $W_{{\rm r},3934}=15-799$
m\AA\ and column densities log $N$(Ca\,{\sc ii}$)=11.25-13.04$ (obtained by
fitting Voigt-profile components).
We obtain a bias-corrected number density of weak intervening Ca\,{\sc ii} absorbers of
$d{\cal N}/dz=0.117\pm0.044$ at $\left<z_{\rm abs}\right> = 0.35$
for absorbers with log $N$(Ca\,{\sc ii}$)\geq 11.65$ ($W_{{\rm r},3934}\geq 32$ m\AA).
This is $\sim 2.6$ times
the value obtained for damped Lyman $\alpha$ absorbers (DLAs) at low redshift.
All Ca\,{\sc ii} absorbers in our sample show associated absorption by other low
ions such as Mg\,{\sc ii} and Fe\,{\sc ii}; 45 percent of them
have associated Na\,{\sc i} absorption.
From ionization modelling we conclude that intervening Ca\,{\sc ii} absorption
with log $N$(Ca\,{\sc ii}$)\geq 11.5$
arises in DLAs, sub-DLAs
and Lyman-limit systems (LLS) at
H\,{\sc i} column densities of log $N$(H\,{\sc i}$)\geq 17.4$.
Using supplementary H\,{\sc i}
information for nine of the absorbers we find that the Ca\,{\sc ii}/H\,{\sc i}
ratio decreases strongly with increasing H\,{\sc i} column density,
indicating a column-density-dependent dust depletion of Ca. The observed
column density distribution function of Ca\,{\sc ii} absorption
components follows a relatively steep power law, $f(N)\propto N^{-\beta}$, with
a slope of $-\beta=-1.68$, which again points towards an enhanced dust depletion
in high column density systems.
The relatively large cross section of these absorbers together with
the frequent detection of Ca\,{\sc ii} absorption in
high-velocity clouds (HVCs) in the halo of the Milky Way suggests
that a considerable fraction of the intervening Ca\,{\sc ii} systems trace
(partly) neutral gas structures in the halos and circumgalactic environment
of galaxies (i.e., they are HVC analogs). Based on the recently measured
detection rate of Ca\,{\sc ii} absorption in the Milky Way HVCs 
we estimate that the mean (projected) Ca\,{\sc ii} covering fraction
of galaxies and their gaseous halos is $\left< f_{\rm c, CaII} \right>=0.33$.
Using this value and considering all galaxies with luminosities
$L\geq0.05L^{\star}$ we calculate that the characteristic radial extent of 
(partly) neutral gas clouds with log $N$(H\,{\sc i}$)\geq 17.4$
around low-redshift galaxies is $R_{\rm HVC}\approx 55$ kpc.
}

%


\titlerunning{Intervening Ca\,{\sc ii} absorbers}

\maketitle

%

\section{Introduction}

The analysis of intervening absorption lines in the
spectra of distant quasars (QSO) has become an extremely
powerful method to study the distribution and physical
properties of the intergalactic medium (IGM) and its
relation to galaxies. Over the last couple of decades, QSO absorption 
spectroscopy of various metal ions such as Mg\,{\sc ii} 
and C\,{\sc iv}, together with galaxy imaging, has been
used extensively 
to constrain the nature of the various processes (infall,
outflow, merging) that govern the matter exchange between
galaxies and the IGM as part of the hierarchical evolution 
of galaxies (e.g., Bergeron \& Boiss\'e 1991; Steidel et al.\,1992;
Churchill et al.\,1999; Steidel et al.\,2002).

Because of the physical and spatial complexity of the 
IGM and the various hydrodynamical processes that are 
involved,
our knowledge about the exact role of the IGM-galaxy
connection for the evolution of galaxies at low and
high redshift is still incomplete. In particular, it
is not known whether the gas infall onto
galaxies occurs in the form of condensed neutral gas 
clouds (``cold'' mode) or in the form of ionized
gas (``warm'' or ``hot'' mode; see, e.g., Bland Hawthorn 2009).
It also remains to be determined
what consequences the accretion mode has for the 
star-formation activity of a galaxy. Part of this 
lack of understanding (particularly at low redshift)
is related to observational limitations.
Since, by far, most of the ion transitions 
of interest for QSO absorption spectroscopy
are located in the ultraviolet (UV),
QSO absorption line observations at low redshift
are limited by the number of bright extragalactic
background sources for which high-resolution UV spectroscopy
with current space-based spectrographs 
(i.e., HST/{\it STIS} or HST/{\it COS}) can be carried out.
At higher redshift, more and more UV lines are redshifted into
the optical regime, allowing us to obtain a much larger
number of high-quality QSO absorption spectra with 
large, ground-based telescopes such as the ESO
{\it Very Large Telescope} (VLT). Because galaxies at
high redshift are dim, however, the characterization
of the IGM-galaxy connection at high $z$ using
direct observations is a challenging task (Adelberger 
et al.\,2003; Crighton et al.\,2010). Fortunately, 
hydrodynamical simulations of the IGM 
have turned out to be an extremely valuable tool to
assess the properties of intergalactic matter and its
relation to galaxies at high and low redshifts 
(e.g., Fangano, Ferrara \& Richter\,2007;
Kaufmann et al.\,2009; Kacprzak et al.\,2010a).

Most of the recent absorption line studies
of intervening systems have concentrated on the analysis of ionic
transitions of hydrogen and heavy elements in the UV 
(e.g, H\,{\sc i}, Mg\,{\sc ii}, C\,{\sc iv}, O\,{\sc vi}, 
and many others). The two most important optical absorption
lines for IGM studies, the Ca\,{\sc ii} H \& K lines near 
4000\,\AA, are less commonly used,
although they have also played an important
role in our understanding of the diffuse matter in space.
With an ionization potential of $11.9$ eV Ca\,{\sc ii}
serves as a tracer of warm and cold, neutral gas in galaxies and
their gaseous halos.
After all, the Ca\,{\sc ii} H \& K lines were the first 
absorption lines ever detected in interstellar gas (Hartmann 1904).
These lines were also the first tracers of extraplanar 
gas clouds in the halo of the Milky Way (M\"unch 1952) and
were among the first metal absorption lines detected in
the IGM in early QSO spectra (see Blades 1988). 

Bowen (1991), Bowen et al.\,(1991), and, more 
recently, Wild, Hewett \& Pettini (2006, 2007), 
Zych et al.\,(2007, 2009) and Nestor et al.\,(2008)
have investigated intervening Ca\,{\sc ii} absorbers
towards QSOs at relatively low spectral resolution.
From the early studies by Bowen et al.\, (using a small sample
of medium resolution optical QSO spectra) it was concluded 
that intervening Ca\,{\sc ii} absorption predominantly occurs within
the inner 20 kpc of galaxies, thus speaking against many galaxies having
extended Ca\,{\sc ii} absorbing neutral gas halos with large filling factors.
From the analysis of several
thousand low-resolution spectra from the Sloan Digital
Sky Survey (SDSS) Wild, Hewett \& Pettini (2006) concluded that
strong Ca\,{\sc ii} systems with rest frame 
equivalent widths $>350$ m\AA\,arise only from the
inner regions of galaxies and thus represent a
subclass of the damped Lyman $\alpha$ absorbers (DLAs). 
These have the highest H\,{\sc i} column densities of 
all intervening QSO absorbers (log $N$(H\,{\sc i}$)\geq 20.3$) 
and are believed to represent gas-rich galaxies and 
protogalactic structures.

However, because of their limited spectral resolution and 
their typically low signal-to-noise ratios (S/N), previous
studies would have 
missed a population of weak, intervening Ca\,{\sc ii} 
absorbers with rest frame equivalent widths $\le 100$ m\AA.
The existence of these
weak Ca\,{\sc ii} systems was recently indicated
by high-resolution observations of weak 
Ca\,{\sc ii} absorption in the halo and 
nearby intergalactic environment of the Milky Way 
(Richter et al.\,2005, 2009; Ben Bekhti et al.\,2008; 
Wakker et al.\,2007, 2008).
It was found that Ca\,{\sc ii} absorption is common in
the many neutral gas structures outside the Milky Way disk (i.e., 
in the so-called ``high velocity clouds'', HVCs),
even at low H\,{\sc i} gas column densities, 
$N$(H\,{\sc i}$)\sim 10^{17}-10^{19}$ cm$^{-2}$. 
If the Milky Way environment is typical 
for low-redshift galaxies, weak Ca\,{\sc ii} absorption 
should arise in both the neutral gas
disks of galaxies {\it and} in their extended, patchy neutral
gas halos (i.e., in HVC analogs), but high sensitivity
is required to detect the halo component.
In terms of the common QSO absorption line classification this implies 
that Ca\,{\sc ii} absorption should arise in DLAs
as well as in a certain fraction of sub-DLAs
($19.3<$log $N$(H\,{\sc i}$)\leq 20.3$) and
Lyman limit systems (LLS; $17.2<$log $N$(H\,{\sc i}$)\leq 19.3$),
depending on the Ca abundance and dust depletion
in the gas and the local ionization conditions. 

The common detection of weak Ca\,{\sc ii} absorption in the 
Milky Way's extended neutral gas halo therefore suggests that the
occurrence of Ca\,{\sc ii} absorption in the gaseous outskirts of 
low-redshift galaxies should be reassessed with
high resolution, high S/N absorption line data. Such a study may be
very useful to constrain the absorption cross section of 
neutral gas in the local Universe. Most importantly, it can
be used to measure the sizes and masses of neutral gas galaxy halos.
The latter point is particularly interesting, as knowledge about
the neutral gas distribution around galaxies {\it below} the H\,{\sc i} 
21cm detection limit of current radio telescopes may be
crucial for our understanding of gas accretion rates of 
galaxies at low redshift (cold accretion vs. warm accretion; 
see Bland Hawthorn 2008).
Because intervening Ca\,{\sc ii} absorbers should trace neutral gas at column
densities log $N$(H\,{\sc i}$)>17$ it is expected that
they are relatively rare.  Their number densities should be lower than the
value of $dN/dz \approx 0.6$ at $z=0.3$, derived for 
the so-called ``strong'' Mg\,{\sc ii} systems 
(systems with rest frame equivalent widths $\geq 0.3$ \AA\,
in the Mg\,{\sc ii} $\lambda 2976$ line; see Nestor, Turnshek
\& Rao 2005), but higher than that of DLAs 
($dN/dz=0.045$ at $z=0$; Zwaan et al.\,2005).  
Therefore, a large number of QSO sightlines must be surveyed in order
to study the properties of intervening Ca\,{\sc ii} systems on a
statistically secure basis.

We here present the first systematic study of weak intervening Ca\,{\sc ii} 
absorbers in the redshift range $z=0-0.5$, based on a very large
sample of more than 300 optical, high-resolution 
($R>40,000$) QSO absorption line spectra 
obtained with the VLT. Our paper is organized as follows. In
Sect.\,2 we present the observations, the data handling, and the
analysis method. The main results of our survey are presented
in Sect.\,3. In Sect.\,4 we discuss the physical properties
and the origin of the weak intervening Ca\,{\sc ii} systems.
We compare our results with previous studies of intervening
Ca\,{\sc ii} absorbers in Sect.\,5.
Finally, we present a summary of our study in Sect.\,6.

%

\begin{figure*}[ht!]
\centering
\resizebox{1.0\hsize}{!}{\includegraphics{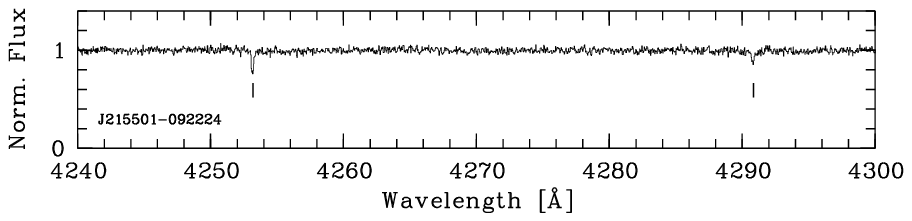}}
\caption[]{
UVES spectrum of the quasar J215501$-$092224 in the wavelength
range between 4240 and 4300 \AA. Weak intervening
Ca\,{\sc ii} absorption is clearly detected at $z=0.08091$ in
both Ca\,{\sc ii} lines, as indicated by the tick marks.
\label{spectrum}
}
\end{figure*}

%

\section{Observations, data handling, and analysis method}

%

\begin{table*}[th!]
\caption[]{QSO names, redshifts and ions of 23 intervening Ca\,{\sc ii} absorbers 
at $z=0-0.5$ detected in our QSO sample}
\label{qsonames}
\begin{normalsize}
\begin{tabular}{llclll}
\hline
QSO & Alt.\,Name & $z_{\rm abs}$ & Bias$^{\rm a}$ & Detected Ions$^{\rm b}$ & Undetected Ions$^{\rm b}$ \\
\hline
J121509+330955 & Ton\,1480      & 0.00396  & yes & Ca\,{\sc ii}, Na\,{\sc i}                               &             \\
J133007-205616 & PKS\,1327-206  & 0.01831  & no  & Ca\,{\sc ii}, Na\,{\sc i}                               &             \\
J215501-092224 & PHL\,1811      & 0.08091  & no  & Ca\,{\sc ii}                                            & Na\,{\sc i} \\
J044117-431343 & HE\,0439-4319  & 0.10114  & yes & Ca\,{\sc ii}, Na\,{\sc i}                               &             \\
J095456+174331 & PKS\,0952+179  & 0.23782  & yes & Ca\,{\sc ii}, Mg\,{\sc ii}, Fe\,{\sc ii}                &             \\
J012517-001828 & PKS\,0122-005  & 0.23864  & no  & Ca\,{\sc ii}, Mg\,{\sc ii}                              &             \\
J235731-112539 & PKS\,2354-117  & 0.24763  & yes & Ca\,{\sc ii}, Na\,{\sc i}, Mg\,{\sc ii}                 &             \\
J000344-232355 & HE\,0001-2340  & 0.27051  & no  & Ca\,{\sc ii}, Mg\,{\sc ii}, Fe\,{\sc ii}                & Na\,{\sc i} \\
J142249-272756 & PKS\,1419-272  & 0.27563  & yes & Ca\,{\sc ii}, Mg\,{\sc ii}, Fe\,{\sc ii}                & Na\,{\sc i} \\
J042707-130253 & PKS\,0424-131  & 0.28929  & no  & Ca\,{\sc ii}, Mg\,{\sc ii}, Fe\,{\sc ii}                &             \\
J113007-144927 & PKS 1127-145   & 0.31273  & yes & Ca\,{\sc ii}, Mg\,{\sc ii}, Fe\,{\sc ii}                &             \\
J102837-010027 & B\,1026-0045B  & 0.32427  & no  & Ca\,{\sc ii}, Mg\,{\sc ii}, Fe\,{\sc ii}                &             \\
J231359-370446 & PKS\,2311-373  & 0.33980  & no  & Ca\,{\sc ii}, Mg\,{\sc ii}, Fe\,{\sc ii}                &             \\
J110325-264515 & PG\,1101-264   & 0.35896  & no  & Ca\,{\sc ii}, Na\,{\sc i}, Mg\,{\sc ii}, Fe\,{\sc ii}   &             \\
J094253-110426 & HE 0940-1050   & 0.39098  & no  & Ca\,{\sc ii}, Mg\,{\sc ii}, Fe\,{\sc ii}                &             \\
J121140+103002 & B\,1209+1046   & 0.39293  & no  & Ca\,{\sc ii}, Mg\,{\sc ii}, Fe\,{\sc ii}                &             \\
J123200-022404 & PKS\,1229-021  & 0.39498  & yes & Ca\,{\sc ii}, Mg\,{\sc ii}, Fe\,{\sc ii}                &             \\
J050112-015914 & PKS\,0458-020  & 0.40310  & no  & Ca\,{\sc ii}, Mg\,{\sc ii}                              & Na\,{\sc i} \\
J224752-123719 & PKS\,2245-128  & 0.40968  & no  & Ca\,{\sc ii}, Mg\,{\sc ii}, Fe\,{\sc ii}                &             \\
J220743-534633 & PKS\,2204-540  & 0.43720  & yes & Ca\,{\sc ii}, Mg\,{\sc ii}, Fe\,{\sc ii}                & Na\,{\sc i} \\
J044117-431343 & HE\,0439-4319  & 0.44075  & no  & Ca\,{\sc ii}, Mg\,{\sc ii}                              & Na\,{\sc i} \\
J144653+011356 & B\,1444+0126   & 0.44402  & no  & Ca\,{\sc ii}, Mg\,{\sc ii}, Fe\,{\sc ii}                &             \\
J045608-215909 & HE\,0454-2203  & 0.47439  & no  & Ca\,{\sc ii}, Fe\,{\sc ii}                              &             \\
\hline
\end{tabular}
\noindent
\\
$^{\rm a}$\,Bias flag indicates whether the observations were targeted observations (see Sect.\,3.2)\\
$^{\rm b}$\,Only absorption by Ca\,{\sc ii}, Mg\,{\sc ii}, Fe\,{\sc ii}, and Na\,{\sc i} is considered in this study
\end{normalsize}
\end{table*}

%

%

\begin{figure}[t!]
\centering
\resizebox{1.0\hsize}{!}{\includegraphics{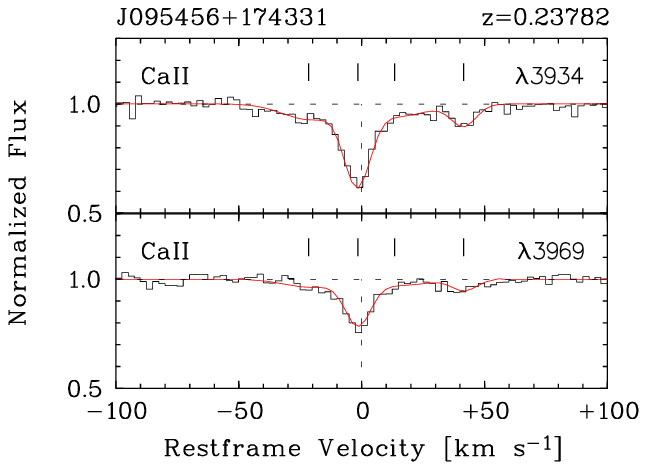}}
\caption[]{Example for a Voigt-profile fit
(red solid line) with four absorption components
of the Ca\,{\sc ii} absorber at
$z=0.23782$ towards J095456+174331.}
\label{voigtfit}
\end{figure}

%

%

\begin{figure}[h!]
\centering
\resizebox{0.9\hsize}{!}{\includegraphics{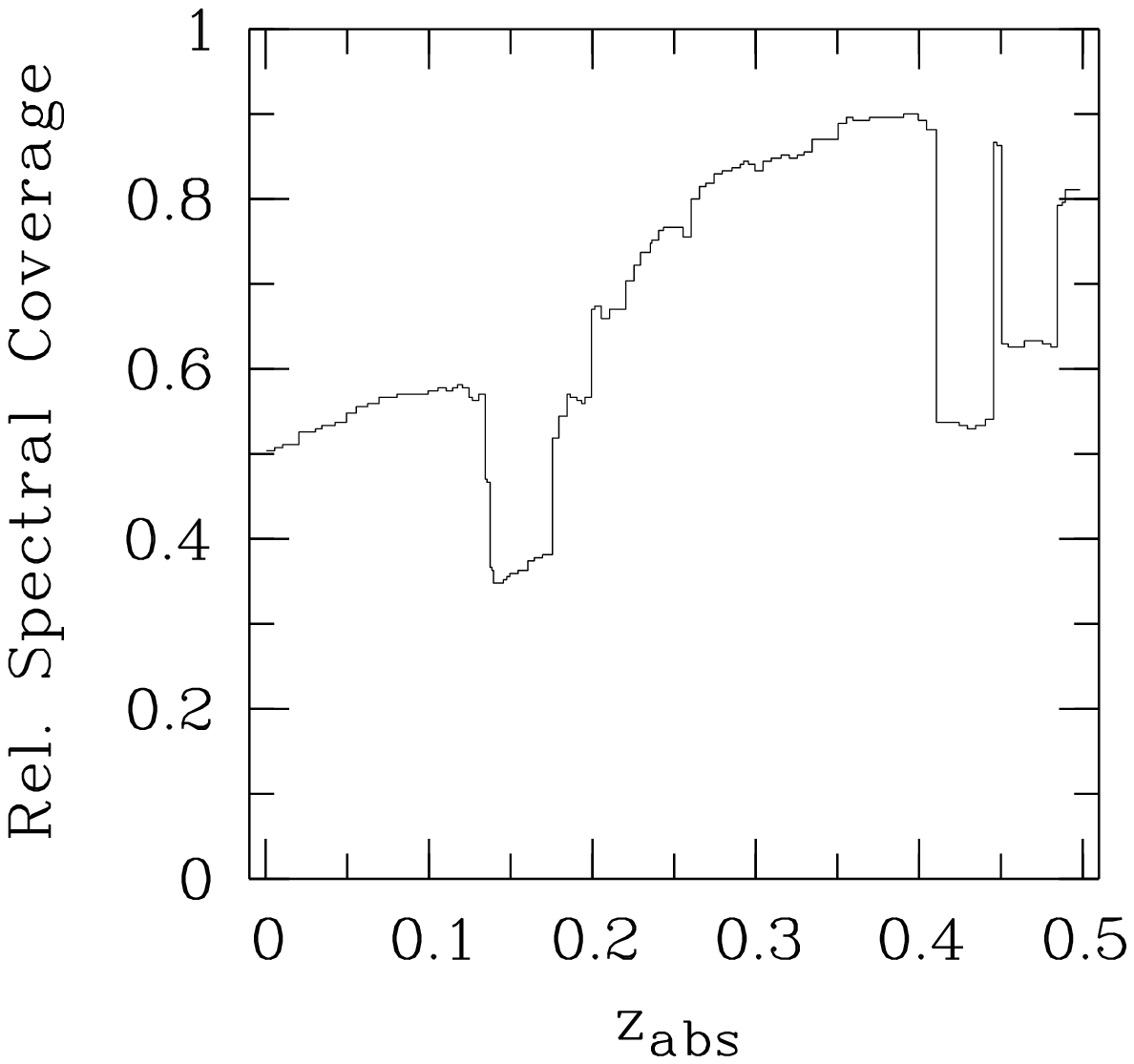}}
\caption[]{Relative redshift coverage for Ca\,{\sc ii}
absorption for 270 QSO sigthlines,
along which intervening Ca\,{\sc ii} absorption
can be detected at column densities
log $N$(Ca\,{\sc ii}$)>11.65$.}
\label{zcoverage}
\end{figure}

%

\subsection{Spectra selection and reduction}

For our Ca\,{\sc ii} survey we made use of the ESO data 
archive\footnote{http://archive.eso.org}
and retrieved all publically available (as of Sept. 2008)
absorption-line data for low- and high-redshift QSOs
observed with the {\it Ultraviolet and Visual Echelle
Spectrograph} (UVES) on the VLT. This enormous 
data archive (Spectral Quasar Absorption Database, 
SQUAD; PI: M.T. Murphy) provides high-quality spectral data 
for $\sim 400$ quasars and active galactic nuclei (AGN).
Most of these spectra were taken in the UVES
standard configuration using the 1'' slit, providing
a spectral resolution of $R\sim45,000$ (corresponding
to a velocity resolution of $\sim 6.6$ km\,s$^{-1}$ FWHM).
Only a few spectra have been observed at slightly higher
spectral resolution, up to $R\sim60,000$. The spectral
coverage as well as the signal-to-noise ratio (S/N) varies
substantially among the spectra, reflecting the various
scientific goals of the original proposals.

All data were reduced using a modified version
of the UVES reduction pipeline. The modifications
were implemented to improve the flux extraction and wavelength
calibration for the different echelle orders. Different
orders then were combined using the custom-written code
{\tt UVES\_popler}
with inverse variance weighting and a cosmic ray rejection
algorithm, to form a single spectrum with a dispersion of
$2.5$ km\,s$^{-1}$ per pixel. 
Finally, the spectra were continuum-normalized with an automated
continuum fitting algorithm.

Because we aim to analyse Ca\,{\sc ii} H\&K ($\lambda\lambda 3934.77,3969.59$) 
absorption in the redshift range $z=0-0.5$, we are interested
in the wavelength range between $3934$ and $5955$~\AA.
From all 397 available quasar spectra in the SQUAD sample
we selected the 304 spectra that a) fully or partly 
cover the above mentioned wavelength range, and b) have 
an average S/N of $\geq 12$ per $\sim 6.6$ km\,s$^{-1}$ wide
resolution element for $R=45,000$; see above).
Note that we require the occurrence of both Ca\,{\sc ii} absorption
lines in the spectrum for a positive detection (see also
Sect.\,2.2). The minimum column density that can be detected
at $4\sigma$ significance for an unresolved, optically 
thin absorption line at wavelength, $\lambda_0$, with
oscillator strength, $f$, in a spectrum with a given
S/N per resolution element and a spectral 
resolution $R=\lambda/\Delta \lambda$ can be
calculated (e.g., Tumlinson et al.\,2002; Richter et al.\,2001) via

%

\begin{equation}
N\geq1.13\times10^{20}\,{\rm cm}^{-2}\,
\frac{4}{R\,{\rm (S/N)}\,f\,(\lambda_0/{\rm \AA)}}.
\end{equation}

%

In our case we have $\lambda_0=3969.59$ \AA\,and
$f=0.3145$ for the weaker of the two Ca\,{\sc ii} 
lines, so that the minimum S/N of 12 in our sample
corresponds to a column density threshold of 
$N$(Ca\,{\sc ii}$)=6.7\times10^{11}$ cm$^{-2}$ 
(log $N$(Ca\,{\sc ii}$)=11.83$) for $R=45,000$.
This column density limit corresponds to an equivalent
width limit of $50$ m\AA\, in the stronger Ca\,{\sc ii}
$\lambda 3934$ line.

For all 304 selected QSO spectra we then
scanned the spectral region of interest 
($3934-5955$~\AA) and excluded from
further analysis those regions
that are contaminated by Ly\,$\alpha$ forest lines
and other spectral features that would 
cause severe blending problems.
As a result, we define for each sightline a 
characteristic redshift path $dz_{\rm CaII}\leq 0.5$ 
(typically composed of several redshift chunks) along which
intervening Ca\,{\sc ii} could be detected.
The mean redshift path per sightline in our sample 
is $\left< dz_{\rm CaII}\right>=0.33$; the total redshift
path is $\Delta z = 100.60$. 

A table listing all selected 304 spectra with the 
QSO names, the average S/N, and $dz_{\rm CaII}$ is
provided in the Appendix (Tables A.4 and A.5).

\subsection{Absorber identification and line fitting}

The next step in our analysis was the search for 
Ca\,{\sc ii} H\&K absorption
at $z=0-0.5$ along the 304 selected QSO sightlines.
First, we used an automated line-finder algorithm
to identify absorption features whose wavelengths
would correspond to combined Ca\,{\sc ii} 
H\&K absorption in 
intervening absorbers at $z\leq0.5$. In this way
we created a candidate list of possible
Ca\,{\sc ii} systems. Secondly, two of our team
(Richter \& Krause) independently analysed all
304 spectra by eye and identified possible
Ca\,{\sc ii} systems. All three candidate 
lists then were merged into one. For each of the candidate 
systems we then checked for associated absorption in 
Mg\,{\sc ii} $\lambda\lambda 2796,2803$, 
Fe\,{\sc ii} $\lambda\lambda 2586,2600$, and
Na\,{\sc i} $\lambda\lambda 5891,5897$ to verify or exclude
the presence of intervening gas absorption at the
redshift indicated by the candidate Ca\,{\sc ii}
absorption. A large number of false detections is expected,
caused by intervening absorption lines that 
coincidently mimic the wavelength pattern expected for
a redshifted Ca\,{\sc ii} doublet. For a positive
detection we thus require that both Ca\,{\sc ii}
lines are clearly detected and that the two lines show
a similar pattern of absorption components and 
similar line shapes.

For a positive detection at redshift $z\geq0.2$ we 
further require that the Ca\,{\sc ii} absorption is accompanied  
by absorption in Mg\,{\sc ii} or Fe\,{\sc ii}.
For all possible Ca\,{\sc ii} systems at $z\geq0.2$ 
a Mg\,{\sc ii} and/or a strong Fe\,{\sc ii} transition 
is covered. The detection of one of these ions 
at the same redshift as Ca\,{\sc ii} proves the
presence of an intervening Ca\,{\sc ii} system,
whereas the significant non-detection of Mg\,{\sc ii}
and Fe\,{\sc ii} indicates a false detection. The
mandatory coexistence of Mg\,{\sc ii} and Fe\,{\sc ii}
with Ca\,{\sc ii} is justified given relative 
abundances of these ions, their ionization 
potentials, and the oscillator strengths of the above 
listed transitions. Ca\,{\sc ii} absorbing clouds
without significant Mg\,{\sc ii} and Fe\,{\sc ii} absorption
are not expected to exist.

The situation is different for redshifts $z<0.2$, where 
we have no coverage of associated Mg\,{\sc ii} and Fe\,{\sc ii} 
absorption in our optical spectra.
Generally, the significant non-detection of 
Na\,{\sc i} in a Ca\,{\sc ii} candidate system 
does not indicate a false Ca\,{\sc ii} detection.
Na\,{\sc i} exists only in relatively dense neutral gas 
and thus is present only in a subset of the Ca\,{\sc ii} systems.
Fortunately, in our particular survey, all but one Ca\,{\sc ii} 
candidate systems at $z<0.2$ for which Na\,{\sc i} was covered in the
optical spectrum, did happen to have Na\,{\sc i} detected.  
Furthermore, the only Ca\,{\sc ii} candidate system at $z<0.2$ 
for which Na\,{\sc i} was not detected (the system towards J215501-092224),
could be confirmed by existing UV absorption line data from 
HST/STIS (Jenkins et al.\,2003).

Finally, as a check to our procedure, and in order to identify
weaker Ca\,{\sc ii} absorbers, we reversed our search and looked
for Ca\,{\sc ii} absorption in (previously identified) strong MgII
absorbers at $z=0.2-0.5$.
In this way, we identified four very weak CaII absorbers,
with $N$(Ca\,{\sc ii})$<11.5$, which we had missed in our direct
search for Ca\,{\sc ii}. We also found a number of Ca\,{\sc ii} 
absorber candidates, for which only one of the two Ca\,{\sc ii} lines 
is seen, e.g., in systems so weak that 
Ca\,{\sc ii}$\lambda 3969$ was not formally detected, 
or in which one member of the Ca\,{\sc ii} doublet
was blended or was not covered in the spectrum (see Jones
et al.\,2010 for an example case). These candidate
systems are not considered any further in our analysis,
but they are listed in the Appendix in Table A.3.  
We moreover excluded all local Ca\,{\sc ii} absorption
features at radial velocities $0-500$ km\,s$^{-1}$,
as these features are caused by neutral gas in the
Milky Way disk and halo (see Ben Bekhti et al.\,2008).

Using the above outlined detection criteria a final list of 
verified Ca\,{\sc ii} absorbers was produced.
As an example we show in Fig.\,1 the UVES spectrum of the
quasar J215501$-$092224 in the wavelength range between
4240 and 4300 \AA. Weak intervening Ca\,{\sc ii} absorption
is clearly detected at $z=0.08091$ in both Ca\,{\sc ii}
lines (indicated by the tick marks).
For the further spectral analysis of the absorption lines
we made use of the {\tt FITLYMAN} package 
implemented in the ESO-MIDAS analysis software
(Fontana \& Ballester 1995). This routine 
uses a $\chi ^2$ minimization algorithm
to derive column densities ($N$) and Doppler parameters
($b$) via multi-component Voigt profile fitting, 
also taking into account the spectral resolution 
of the instrument. An example for a multicomponent
Voigt profile fit of an intervening Ca\,{\sc ii}
absorber is shown in Fig.\,2. Total 
Ca\,{\sc ii} column densities for each system
have been determined by summing over the 
the column densities of the individual subcomponents.
The total rest frame equivalent 
widths ($W_{\rm r}$) and velocity widths of the absorption 
($\Delta v$) were derived for 
Ca\,{\sc ii} $\lambda 3934$, as well as for
associated Mg\,{\sc ii} $\lambda 2796$ absorption, by 
a direct pixel integration.

%

\section{Results}

%

\begin{table*}[th!]
\caption[]{Column densities, absorption components, velocity widths
and equivalent widths for the 23 intervening Ca\,{\sc ii} absorbers}
\label{columndens}
\begin{small}
\begin{tabular}{lcccrrrrrc}
\hline
QSO & $z_{\rm abs}$ & log $N$(Ca\,{\sc ii}) & $n_{\rm c}^{\rm a}$\ & $\Delta v_{\rm CaII}$ & $W_{\rm r,CaII,3934}$ &
$\Delta v_{\rm MgII}$ & $W_{\rm r,MgII,2796}$ & log $N$(Na\,{\sc i}) \\
 & & ($N$ in [cm$^{-2}$]) & & [km\,s$^{-1}]$ & [m\AA] & [km\,s$^{-1}]$ & [m\AA] & ($N$ in [cm$^{-2}$]) & \\
\hline
J121509+330955 & 0.00396  & 12.31$\pm$0.01 & 3 & 149 & 160$\pm$9  &       ... &         ...       & 11.64$\pm$0.04 \\
J133007-205616 & 0.01831  & 13.04$\pm$0.06 & 9 & 378 & 799$\pm$46 &       ... &        ...        & 13.28$\pm$0.05 \\
J215501-092224 & 0.08091  & 11.76$\pm$0.02 & 1 &  38 &  44$\pm$5  &       ... &        ...        & $\leq 10.49$   \\
J044117-431343 & 0.10114  & 12.60$\pm$0.05 & 7 & 130 & 292$\pm$6  &       ... &       ...         & 12.25$\pm$0.03 \\
J095456+174331 & 0.23782  & 12.22$\pm$0.07 & 4 & 105 & 119$\pm$7  &       173 &       1101$\pm$12 & ... & ...\\
J012517-001828 & 0.23864  & 11.48$\pm$0.06 & 1 &  22 &  34$\pm$4  &       154 &        328$\pm$30 & ... & ...\\
J235731-112539 & 0.24763  & 12.70$\pm$0.03 & 3 &  53 & 261$\pm$16 & $\geq$470 & $\geq$2244        & 12.41$\pm$0.02 & \\
J000344-232355 & 0.27051  & 11.66$\pm$0.02 & 2 &  42 &  33$\pm$2  &       176 &       850$\pm$7   & $\leq 10.30$ \\
J142249-272756 & 0.27563  & 12.07$\pm$0.03 & 1 &  29 &  80$\pm$6  &       302 &       1260$\pm$50 & $\leq 10.90$ \\
J042707-130253 & 0.28929  & 12.03$\pm$0.03 & 2 & 122 &  67$\pm$2  &       158 &        368$\pm$13 & ... & ...\\
J113007-144927 & 0.31273  & 12.71$\pm$0.01 & 8 & 230 & 376$\pm$7  &       378 &       1789$\pm$12 & ... & ...\\
J102837-010027 & 0.32427  & 12.43$\pm$0.02 & 2 & 143 & 224$\pm$14 &       165 &        681$\pm$20 & ... & ...\\
J231359-370446 & 0.33980  & 12.81$\pm$0.04 & 3 &  41 & 166$\pm$5  &       149 &        976$\pm$16 & ... & ...\\
J110325-264515 & 0.35896  & 11.26$\pm$0.02 & 1 &  31 &  15$\pm$2  &       148 &        532$\pm$15 & 11.92$\pm$0.01 \\
J094253-110426 & 0.39098  & 11.89$\pm$0.02 & 2 &  31 &  54$\pm$2  & $\geq$178 & $\geq$900         & ... & ...\\
J121140+103002 & 0.39293  & 11.38$\pm$0.06 & 1 &  30 &  18$\pm$3  &       355 &       1114$\pm$15 & ... & ...\\
J123200-022404 & 0.39498  & 12.39$\pm$0.05 & 5 & 111 & 194$\pm$6  &       297 &       2018$\pm$10 & ... & ...\\
J050112-015914 & 0.40310  & 12.26$\pm$0.03 & 3 & 103 & 132$\pm$17 &       ... & $\geq$715         & $\leq 11.20$ \\
J224752-123719 & 0.40968  & 12.23$\pm$0.02 & 2 & 101 & 140$\pm$11 &       448 &       1013$\pm$21 & ... & ...\\
J220743-534633 & 0.43720  & 11.98$\pm$0.05 & 1 &  26 &  65$\pm$8  &        68 &        318$\pm$15 & $\leq 11.30$ \\
J044117-431343 & 0.44075  & 11.42$\pm$0.07 & 3 &  60 &  24$\pm$3  &        72 &        317$\pm$11 & $\leq 10.80$ \\
J144653+011356 & 0.44402  & 11.25$\pm$0.06 & 1 &  42 &  21$\pm$6  &       188 &        219$\pm$10 & ... & ...\\
J045608-215909 & 0.47439  & 12.18$\pm$0.07 & 4 & 102 & 122$\pm$5  &       ... &        ...        & ... & ...\\
\hline
\end{tabular}
\noindent
\\
$^{\rm a}$\,$n_{\rm c}$ is the number of Ca\,{\sc ii} absorption components per system; 
see Tables A.1 and A.2\\

\end{small}
\end{table*}

%

\subsection{Frequency and column densities of Ca\,{\sc ii} absorbers}

Based on the analysis method described in the previous
section we found 23 intervening Ca\,{\sc ii} absorbers
along the selected 304 QSO sightlines. The redshift of the
absorbers ranges between $z=0.00396$ and $z=0.47439$.
Note that several of these absorption systems represent DLAs and 
LLSs that have been identified and studied previously,
based on the analysis of ions other than Ca\,{\sc ii} (e.g., 
Mg\,{\sc ii}). Consequently, it is expected that some of our 
QSO sightlines that exhibit intervening Ca\,{\sc ii} absorption
have been explicitly selected (based on lower resolution spectra)
to study the absorption 
characteristics of DLAs and LLSs at low $z$, and thus represent
targeted observations. Therefore, one important concern about 
the detection rate of intervening Ca\,{\sc ii} in our data 
is the {\it selection bias} in the QSO sample that we are using.
This issue will be further discussed in Sect.\,3.2, where we
estimate the number density of Ca\,{\sc ii} absorbers at low
redshift.

Quasar names, redshifts and detected (and undetected) ions
for the 23 Ca\,{\sc ii} absorption 
systems are listed in Table 1. Detailed
results from the Voigt-profile fitting of the absorbers
and their subcomponents are listed in Tables A.1 and A.2 in
the Appendix. Velocity plots of Ca\,{\sc ii} and other
detected ions for all 23 absorbers are shown in the 
Appendix, in Figs.\,A.1 to A.6.
Total column densities, absorption components, velocity widths,
and equivalent widths for the 23 Ca\,{\sc ii} absorbers and
for their associated Mg\,{\sc ii} absorption are given in Table 2.

The total logarithmic Ca\,{\sc ii} column densities of the 
absorbers range between log $N$(Ca\,{\sc ii}$)=11.25$ and
$13.04$. The distribution of the (total) 
logarithmic Ca\,{\sc ii} column 
densities of the 23 absorbers as well as the distribution 
of their redshifts are shown in Fig.\,4. The
Ca\,{\sc ii} column density distribution has  
a peak near log $N$(Ca\,{\sc ii}$)\sim 12.2$, which
is also the median value. If we remove the eight systems
that are flagged as ``biased'' in Table 1, the median
value for log $N$ is reduced to $\sim 11.9$ (for details 
on the selection bias see Sect.\,3.2).
The redshift distribution of the absorbers 
(Fig.\,4, right panel) is nonuniform, showing
an enhancement of Ca\,{\sc ii} systems in the range
$z=0.2-0.4$. This is because of the inhomogeneous
redshift coverage of our QSO sample and not because of
an intrinsic redshift evolution of the absorber
population (see Sect.\,3.2).
The median absorption redshift in our total sample 
(23 systems) is $z_{\rm abs}=0.32$ ($0.35$ 
in the bias-corrected sample).

%

\begin{figure*}[t!]
\centering
\resizebox{0.8\hsize}{!}{\includegraphics{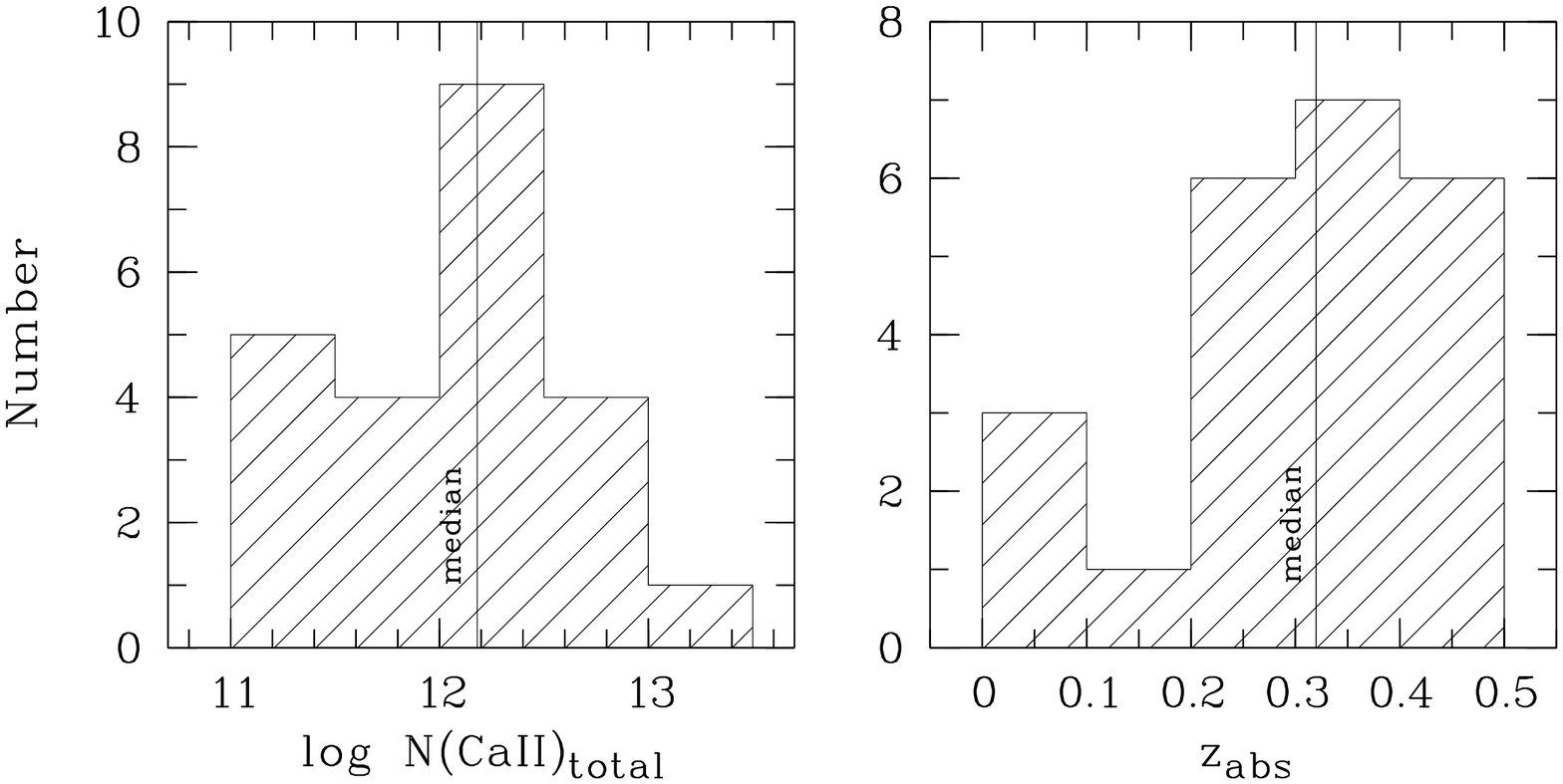}}
\caption[]{Distribution of (total) logarithmic
Ca\,{\sc ii} column densities of the 23 absorbers
detected in our total QSO sample (left panel) and
their redshift distribution
(right panel).}
\label{columndensities}
\end{figure*}

%

\begin{figure*}[t!]
\centering
\resizebox{0.8\hsize}{!}{\includegraphics{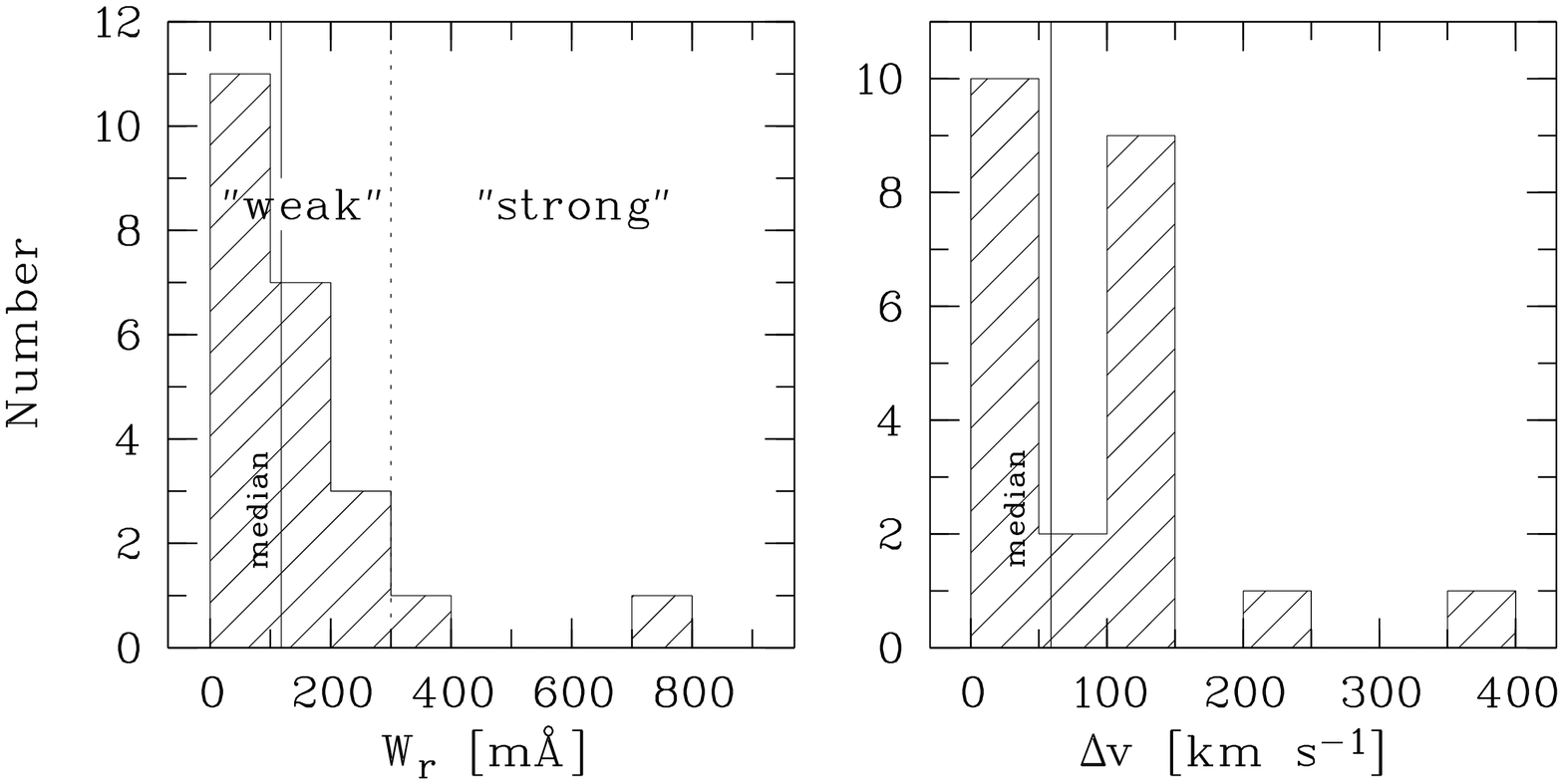}}
\caption[]{Distribution of Ca\,{\sc ii} $\lambda 3934$
rest frame equivalent widths for the 23
Ca\,{\sc ii} absorbers (left panel)
and their velocity width distribution (right panel).}
\label{ewdist}
\end{figure*}

%

\begin{figure*}[t!]
\centering
\resizebox{0.8\hsize}{!}{\includegraphics{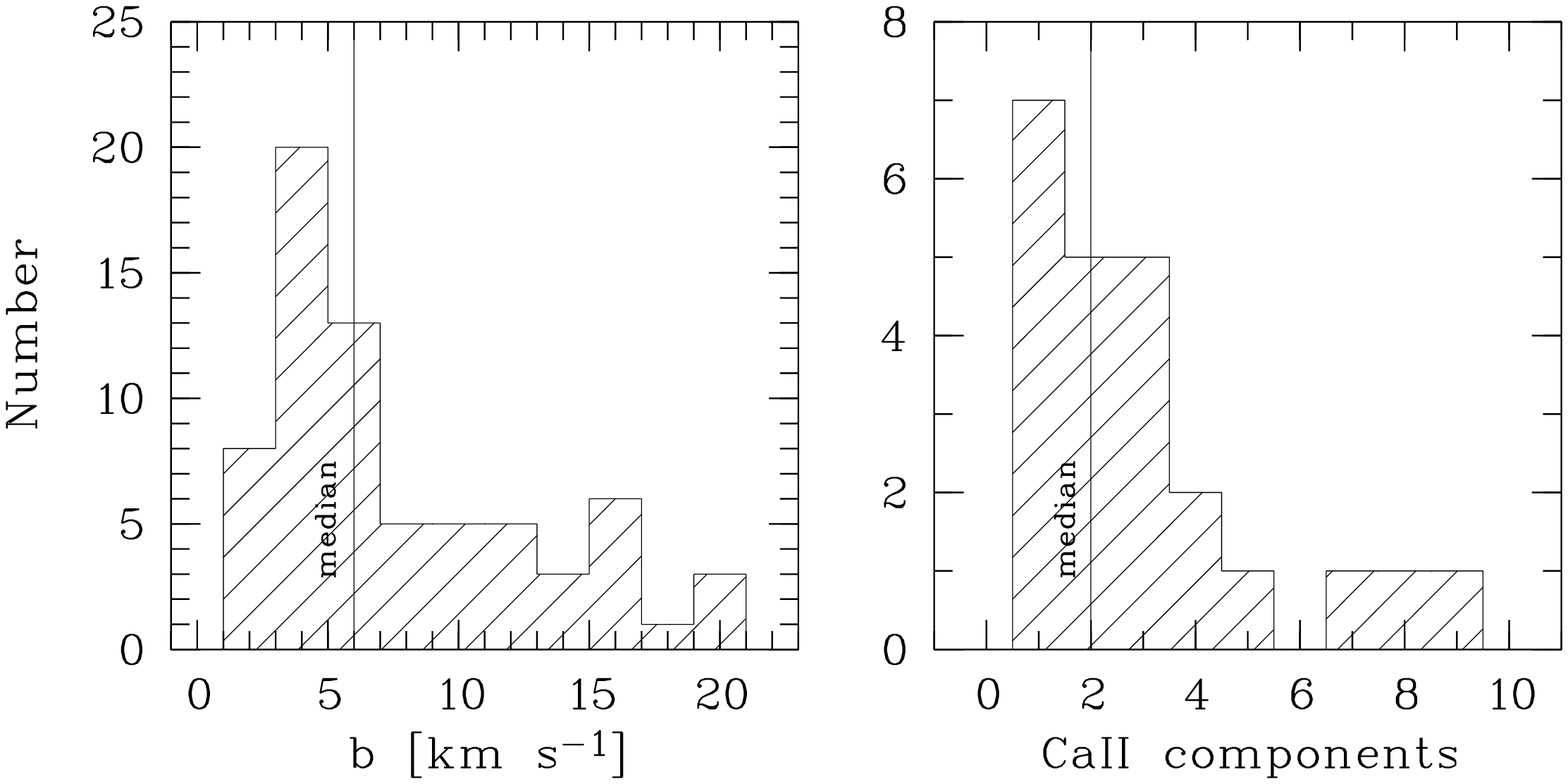}}
\caption[]{Distribution of Doppler parameters ($b$ values)
of the 69 Ca\,{\sc ii} absorption components in our
23 Ca\,{\sc ii} systems (left panel), and number
of absorption components per system (right panel), derived
from Voigt-profile fitting.
}
\label{bdist}
\end{figure*}

%

\subsection{Number density of Ca\,{\sc ii} absorbers}

To calculate the number density of intervening
Ca\,{\sc ii} absorbers per unit redshifz, $d{\cal N}/dz$,
it is necessary to consider in detail the completeness of our
Ca\,{\sc ii} survey and the selection bias in our
data sample.

The weakest Ca\,{\sc ii} system in our survey
has log $N$(Ca\,{\sc ii}$)=11.25$,
but only 103 of the 304 spectra have a sufficiently high S/N to
detect Ca\,{\sc ii} at this column density level (see
Eq.\,1). Eighteen Ca\,{\sc ii} absorbers have
log $N$(Ca\,{\sc ii}$)>11.65$ (these were found by the direct 
search for intervening Ca\,{\sc ii} absorption; see Sect.\,2.2)
and for 270 out of 304 sightlines the selected spectral regions 
are sensitive above this column density
level. For completeness reasons we therefore consider
only the total redshift path covered by these 
270 sightlines, together with the 18 absorbers that have
log $N$(Ca\,{\sc ii}$)>11.65$, for the estimate of the
Ca\,{\sc ii} absorber number density. 

The relative redshift coverage, $f_z(z)$, for Ca\,{\sc ii} absorption 
along the selected 270 sightlines ($f_z(z)=\sum dz(z)/270$) 
is shown in Fig.\,3.
The distribution reflects blending issues and low S/N
for many high-redshift sightlines in the
range $z_{\rm CaII}<0.25$, as well as individual
features caused by the integrated wavelength coverage
of the original UVES observations in our sample.
The total redshift path for the selected 270
sightlines is $\Delta z=89.15$.

To quantify the selection bias in our data we inspected
the original proposal abstracts in the UVES archive for the
sightlines along which intervening Ca\,{\sc ii} was found.
As it turns out, 8 out of the 18 QSO spectra with intervening
Ca\,{\sc ii} absorption at log $N$(Ca\,{\sc ii}$)>11.65$
represent data from targeted observations
of DLAs and Mg\,{\sc ii} absorbers at low redshift
(as indicated in Table 1, Col.\,4). This implies that our sample
contains $\sim 40$ percent more Ca\,{\sc ii} systems per
unit redshift than in a fully random QSO sample. 

Another way to check for a possible selection bias in our
data is to compare the frequency of Mg\,{\sc ii} absorption
systems in our sample with that of securely unbiased absorber
searches (e.g., from spectra of the Sloan Digital Sky
Survey, SDSS). 
Intervening Mg\,{\sc ii} absorbers commonly are divided into
``weak'' systems (for rest-frame equivalent widths in the
$\lambda 2976$ line, $W_{2796}\leq 300$ m\AA) and ``strong''
systems ($W_{2796}\geq 300$ m\AA). 
From a Mg\,{\sc ii}-selected absorber search of our data 
we find that strong Mg\,{\sc ii} systems outnumber 
Ca\,{\sc ii} systems with log $N$(Ca\,{\sc ii}$)>11.65$
by a factor of $\sim 5$. Moreover, Ca\,{\sc ii} systems
above a column density limit of log $N=11.65$ arise
exclusively in strong Mg\,{\sc ii} systems, but not
in weak Mg\,{\sc ii} absorbers. Therefore, the number
densities of strong Mg\,{\sc ii} absorbers and 
Ca\,{\sc ii} systems with log $N$(Ca\,{\sc ii}$)>11.65$
can be directly compared to each other.
From our data we estimate $d{\cal N}/dz\approx 1.0$ for strong 
Mg\,{\sc ii} systems at $\left< z \right> = 0.3$. 
The study by Nestor, Turnshek \& Rao (2005)
based on SDSS data, however, indicates a 
lower value of $d{\cal N}/dz\approx 0.6$
for this redshift. These numbers further suggest a 
substantial, $\sim 40$ percent overabundance 
of strong Mg\,{\sc ii} systems, and thus Ca\,{\sc ii} systems
with log $N$(Ca\,{\sc ii}$)>11.65$ in our 
QSO sample, owing to a selection bias.

Without any bias correction our detection of 18
Ca\,{\sc ii} absorbers with log $N$(Ca\,{\sc ii}$)>11.65$
along a total redshift path of $\Delta z=89.15$ implies 
$d{\cal N}/dz$(Ca\,{\sc ii}$)=0.202\pm0.048$ for this
column density range.
Considering the above estimate for the 
selection bias (40 percent) and removing the
eight relevant sightlines from the statistics we derive 
a bias-corrected number density of 
intervening Ca\,{\sc ii} systems with 
log $N$(Ca\,{\sc ii}$)>11.65$ of
$d{\cal N}/dz$(Ca\,{\sc ii}$)=0.117\pm0.044$. 
The error includes a 20 percent uncertainty
estimate for the bias correction.
For comparison, the number density of DLAs at
low redshift is $d{\cal N}/dz$(DLA$)\approx0.045$ 
(Zwaan et al.\,2005), thus a factor of 2.6 lower.
This is an interesting and important result: {\it weak 
intervening Ca\,{\sc ii} absorbers at low redshift 
outnumber DLAs by a factor of two to three}.

If we transform $dz$ into the comoving absorption path
length $dX$ via

%

\begin{equation}
dX=(1+z)^2\,[\Omega_{\Lambda}+\Omega_{\rm m}(1+z)^3]^{-0.5}\,dz,
\end{equation}

%

the data indicate $d{\cal N}/dX=0.075$ for $\left< z_{\rm abs} \right> =0.35$ 
(bias corrected).
Thoughout this paper we use a flat $\Lambda$CDM cosmology 
with $H_{\rm 0}=73$ km\,s$^{-1}$\,Mpc$^{-1}$, 
$\Omega_{\rm m}=0.238$, and $\Omega_{\Lambda}=0.762$ (Spergel et al.\,2007).

%

\begin{figure}[t!]
\centering
\resizebox{1.0\hsize}{!}{\includegraphics{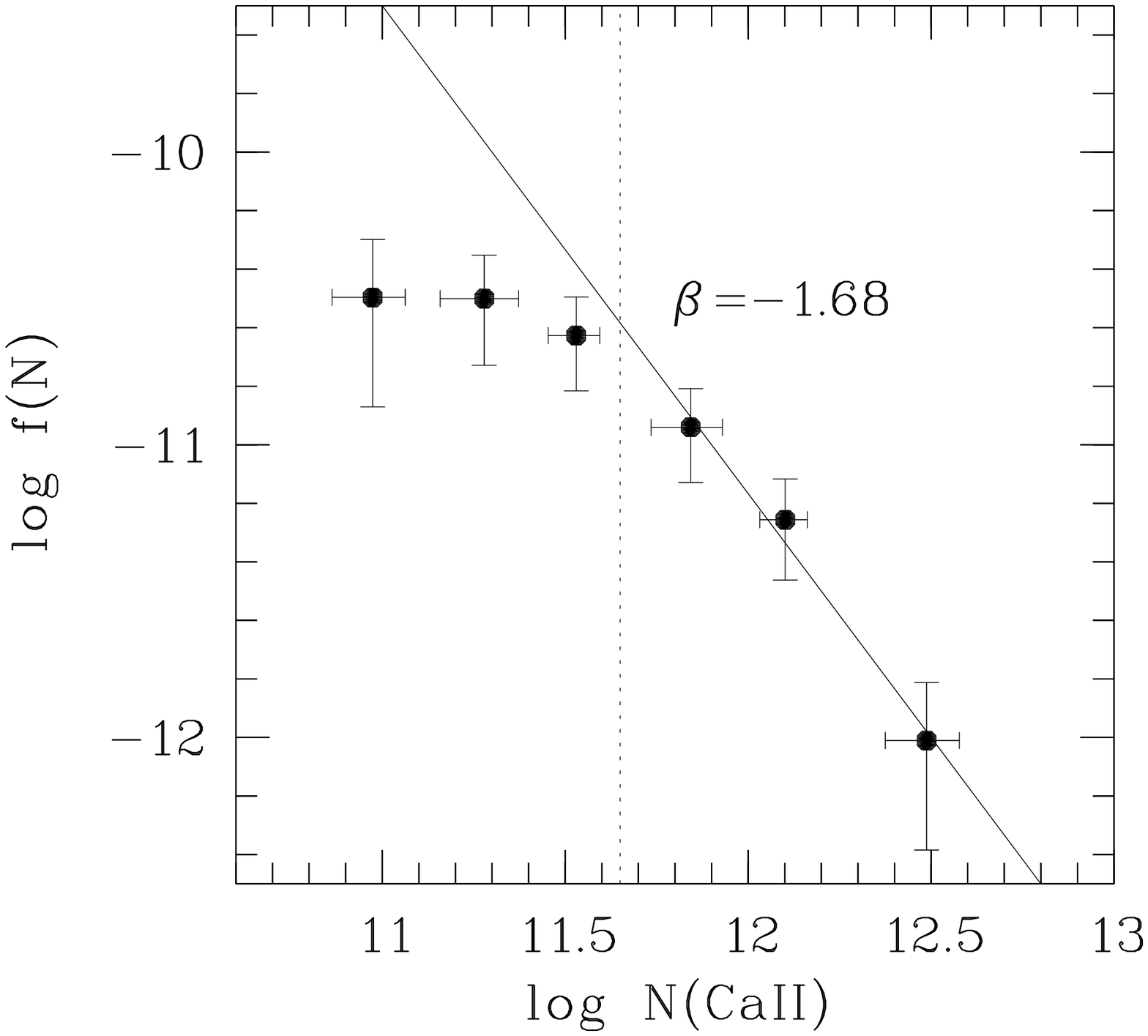}}
\caption[]{Column density distribution function, $f(N)$,
for Ca\,{\sc ii} components,
as derived from our bias-corrected Ca\,{\sc ii} absorber sample. Above our
completeness limit (log $N=11.65$; dotted line), the data fit best
to a power law of the form $f(N)=C\,N^{-\beta}$ with
slope $\beta = 1.68 \pm 0.20$ (solid line).}
\label{cddf}
\end{figure}

%

\subsection{Ca\,{\sc ii} equivalent and velocity widths}

The Ca\,{\sc ii} rest-frame equivalent widths for 
the stronger $\lambda 3934$ transition, $W_{\rm r,3934}$,
range from $15$ to $799$ m\AA\, in our sample (see Table 2).
The equivalent width distribution is shown in Fig.\,5,
left panel. Most absorbers (18 out of 23) have 
$W_{\rm r}<200$ m\AA; note that these systems 
would be invisible in spectral data with low 
spectral resolution and/or low S/N (e.g., in most SDSS spectra).
The median rest-frame equivalent width in the
$\lambda 3934$ line is $118$ m\AA\,
($76$ m\AA\,in the unbiased sample).
If we divide our absorber sample into ``strong''
absorbers with $W_{\rm r}\geq 300$ m\AA\, and 
``weak'' absorbers with $W_{\rm r}< 300$ m\AA\,
(similar to the divisions for Mg\,{\sc ii}), our sample 
contains 21 weak absorbers, but only two 
strong systems.
Our study thus indicates that weak Ca\,{\sc ii} 
absorbers with $32 \leq W_{\rm r,3934} < 300$ m\AA
\footnote{$32$ m\AA\,corresond to our completeness
level of log $N=11.65$ for optically thin Ca\,{\sc ii} 
absorption in an unresolved, single-component 
absorption line.} are eight times more numerous than
strong absorbers.

Figure 5, right panel, shows the distribution of 
the absorption width of the Ca\,{\sc ii} systems,
$\Delta v_{\rm abs}$, as derived from the Ca\,{\sc ii} 
absorption profiles. The absorption width is defined as 
the total velocity range in which Ca\,{\sc ii} 
$\lambda 3934$ absorption 
(including all velocity subcomponents) is 
detected in a system. The Ca\,{\sc ii} absorption widths 
range from $22$ to $378$ km\,s$^{-1}$, but 
only two of the 23 systems show large widths
with $\Delta v>200$ km\,s$^{-1}$.
For $\Delta v<200$ km\,s$^{-1}$ the distribution 
of the 21 systems is bimodal with 10 narrow absorbers
with $\Delta v<50$ km\,s$^{-1}$, and 9 somewhat
broader systems with velocity widths between
$100$ and $150$ km\,s$^{-1}$. 
The median value for the velocity width
of Ca\,{\sc ii} is $60$ km\,s$^{-1}$
($40$ km\,s$^{-1}$ for the unbiased sample). While the
typical velocity spread of intervening Ca\,{\sc ii}
absorbers is $<150$ km\,s$^{-1}$, the associated Mg\,{\sc ii}
covers an absorption range that typically is
several times larger than that for Ca\,{\sc ii}. This
will be further discussed in Sect.\,3.6.

%

\subsection{Doppler parameters and subcomponent structure}

The distribution of the Doppler parameters of all
individual 69 absorber subcomponents
is shown in the left panel of 
Fig.\,6. The distribution
peaks at $b=4$ km\,s$^{-1}$ with a broad tail that
extends to $b=20$ km\,s$^{-1}$. The median value
is $b=6$ km\,s$^{-1}$. In absorption spectroscopy 
of extended gaseous structures it is usually 
assumed that the total (=measured) Doppler parameter
of an absorber is composed of a thermal component
($b_{\rm th}$) and a non-thermal component ($b_{\rm turb}$),
so that $b^2=b_{\rm th}^2+b_{\rm turb}^2$. 
The thermal component of $b$ depends on the gas
temperature, $T$, and the mass ($m$) or
atomic weight ($A$) of the particle, where
$b_{\rm th}=(2kT/m)^{1/2} \approx 0.129\,(T$[K]$/A)^{1/2}$ 
km\,s$^{-1}$. The non-thermal component includes macroscopic 
motions in the gas, such as turbulence and flows. 
Assuming pure thermal broadening (i.e., $b_{\rm turb}=0$ km\,s$^{-1}$),
Ca\,{\sc ii} Doppler parameters of ($5,10,15,20$) km\,s$^{-1}$
therefore correspond to logarithmic gas temperatures, log ($T$/K),
of ($4.8,5.4,5.7,6.0$).
Since the temperature of the Ca\,{\sc ii} absorbing 
gas is likely to be much lower than that 
(a characteristic temperature range is
$T=10^2-10^4$ K; see Richter et al.\,2005; 
Ben Bekhti et al.\,2008),
this would suggest that the measured Ca\,{\sc ii} $b$ values 
are dominated by turbulent motions in the gas. 
Alternatively, the observed non-thermal line widths 
may be partly due to unresolved subcomponent structure
in the lines, which is plausible in view of the 
limited spectral resolution of the data 
(FWHM$\approx 6.6$ km\,s$^{-1}$ for most of the spectra).
In either case, the measured Doppler parameters of the 
Ca\,{\sc ii} absorption components unfortunately cannot be used 
to constrain the gas temperature in the absorbers. 

Figure 6, right panel, shows the number of absorption 
components that we resolved in each Ca\,{\sc ii} system.
Seventeen out of the 23 Ca\,{\sc ii} absorbers (i.e., 74 percent) 
have three or fewer absorption components within a
velocity range of $\leq 150$ km\,s$^{-1}$,
and seven systems are single-component Ca\,{\sc ii} absorbers.
The mean velocity separation between Ca\,{\sc ii} absorption
sub-components is only $\sim 25$ km\,s$^{-1}$, which 
suggests that the individual subcomponents are physically
connected (e.g., as part of a coherent gas structure).
Note that the velocity width of an absorber (see above) 
is correlated with the the number of absorption components
(see Table 3).

%

\subsection{Column density distribution function (CDDF)}

In Fig.\,7 we show the column density distribution
function (CDDF) of the 37 Ca\,{\sc ii} absorption 
components found in the unbiased absorber sample. 
Following Churchill et al.\,(2003), the
CDDF can be written as $f(N)=m/\Delta N$, where $m$
denotes the number of absorbers in the column density
bin $\Delta N$. Integration of $f(N)$ over the 
total column density range of interest then delivers
the total number of absorbers in that range. The 
CDDF of low and high ion absorbers in intervening
absorption line systems usually follows a 
power law in the form 

%

\begin{equation}
f(N)=C\,N^{-\beta},
\end{equation}

%

where $\beta$ varies in the range between $1$ and $2$ for
different ions (e.g., $\beta\approx 1.5$ for H\,{\sc i}
and Mg\,{\sc ii}; Kim, Christiani \& D'Odorico\,2001; Churchill
et al.\,2003). Fitting $f(N)$ to the 37 Ca\,{\sc ii}
absorption components we find $\beta = 1.68 \pm 0.20$
and log $C=9.05\pm 2.42$ for all absorbers that 
are above our completeness limit (log $N($Ca\,{\sc ii}$)=
11.65$). Thus, the CDDF of intervening Ca\,{\sc ii} 
absorption components is mildly
steeper than that of intervening H\,{\sc i} and 
Mg\,{\sc ii}. Note that a similar slope 
($\beta = 1.6 \pm 0.3$) has been found for the
Ca\,{\sc ii} CDDF in HVCs in the halo of the 
Milky Way (Ben Bekhti et al.\,2008). 

The relatively steep slope of the 
Ca\,{\sc ii} CDDF most likely is related to the depletion
of Ca into dust grains. Dust depletion
is particularly important for high column 
density systems (see Sects.\,3.6 and 4.2). This effect
leads to a reduction of high column density 
Ca\,{\sc ii} systems and thus is expected to cause a 
steepening of $f(N)$ compared to undepleted elements.
Note that if we fit the CDDF to all 69 absorption components
in the total data set (i.e., including the biased absorbers
and their components), the slope of the CDDF steepens substantially 
with $\beta = 2.12 \pm 0.17$. Obviously, the biased systems
predominantly add components with relatively low Ca\,{\sc ii} column
densities to the total sample, resulting in a steeper CDDF.
This most likely is related to an enhanced dust depletion 
in the biased systems.

%

\subsection{Associated Mg\,{\sc ii}, Fe\,{\sc ii} and Na\,{\sc i} absorption}

All Ca\,{\sc ii} absorbers in our sample for which 
information on Mg\,{\sc ii} and Fe\,{\sc ii}
is available (Mg\,{\sc ii}: 18/23, Fe\,{\sc ii}: 15/23;
see Table 1 and Figs.\,A1-A5) do show associated 
absorption in these ions, following our original
selection criteria (Sect.\,2.2). The equivalent widths of 
the associated Mg\,{\sc ii} $\lambda\lambda 2796,2803$ and
Fe\,{\sc ii} $\lambda\lambda 2586,2600$ absorption lines
are typically much larger than those
of the Ca\,{\sc ii} H\&K absorption. This is expected, 
since Mg is 17 times and Fe is 14 times more abundant than Ca 
(assuming relative solar abundances for these elements:
log (Ca/H)$_{\sun}=-5.69\pm0.04$; 
log (Na/H)$_{\sun}=-5.83\pm0.04$;
log (Mg/H)$_{\sun}=-4.47\pm0.09$; 
log (Fe/H)$_{\sun}=-4.55\pm0.05$; 
Asplund, Grevesse \& Sauval 2005) and the oscillator strengths of
the above mentioned Mg\,{\sc ii} and Fe\,{\sc ii} transitions 
are relatively strong. 
Mg\,{\sc ii} and Fe\,{\sc ii} absorption 
that is associated with Ca\,{\sc ii} systems
shows a velocity-component structure that typically 
is far more complex than that of Ca\,{\sc ii}.
The median velocity width of Mg\,{\sc ii} is 
$\Delta v= 175$ km\,s$^{-1}$, thus substantially
larger than the median velocity width of 
Ca\,{\sc ii} ($60$ km\,s$^{-1}$; see
above).
This behaviour indicates that 
Mg\,{\sc ii} absorption traces more extended gaseous 
structures that have a larger velocity extent, while 
Ca\,{\sc ii} absorption is detectable only in certain, 
spatially more confined 
regions within these structures. This is because
Mg\,{\sc ii} 
- in view of its larger abundance and higher
ionization potential - is a much more sensitive 
tracer for diffuse neutral gas than Ca\,{\sc ii},
and it also traces diffuse ionized gas (see
Sect.\,4.1). Therefore, the velocity pattern
of Mg\,{\sc ii} absorption reflects the overall
distribution of dense and diffuse gas in the
absorber, while Ca\,{\sc ii} absorption arises 
only in the regions that have the highest neutral 
gas column densities.

All but one of the Ca\,{\sc ii} absorbers in our sample, for which
information on Mg\,{\sc ii} is available, are associated
with strong Mg\,{\sc ii} absorbers. However, from our Mg\,{\sc ii}
selected absorber sample, we find that only
every fifth strong Mg\,{\sc ii} system at $z\leq 0.5$
shows Ca\,{\sc ii} absorption above log $N$(Ca\,{\sc ii}$)=11.65$
(see Sect.\,3.2).
The median rest-frame 
equivalent width of the Mg\,{\sc ii} $\lambda 2976$ 
absorption that is associated with a Ca\,{\sc ii} system
is $700$ m\AA, thus about six times 
larger than the median Ca\,{\sc ii} equivalent width 
in the $\lambda 3934$ line. 
As shown in Fig.\,8 (left panel), the Mg\,{\sc ii} $\lambda 2976$ 
equivalent width correlates with the Ca\,{\sc ii} 
$\lambda 3934$ equivalent width, while the velocity widths
of the Ca\,{\sc ii} and Mg\,{\sc ii} absorption do not
show such a correlation (Fig.\,8, right panel). The former
aspect indicates that the Ca\,{\sc ii} absorption follows only
the strongest Mg\,{\sc ii} absorption components, which
dominate the total Mg\,{\sc ii} equivalent width.
This is in line with the observation that there is no
obvious velocity offset between the main Ca\,{\sc ii}
and the main Mg\,{\sc ii} absorption components.
The latter aspect shows that weaker Mg\,{\sc ii} satellite
components are either not traced by Ca\,{\sc ii}
(as they may arise in predominantly ionized
gas rather than in neutral gas) or they are
too weak to detect in Ca\,{\sc ii} absorption.

For 11 of the 23 absorbers, information on Na\,{\sc i}
is available, and five of these systems are indeed 
detected in Na\,{\sc i} absorption. The 
Na\,{\sc i} column densities in the systems where
Na\,{\sc i} is detected range between 
log $N$(Na\,{\sc i}$)=11.64$ and $13.28$ (see Table 2).
The upper limits for log $N$(Na\,{\sc i}) for the 
remaining six systems vary between $10.30$ and $11.20$.
Unlike the associated Mg\,{\sc ii} and Fe\,{\sc ii} absorption,
the observed Na\,{\sc i} component structure in the 
five systems that show Na\,{\sc i} absorption 
follows that of the Ca\,{\sc ii} 
absorption very closely (see Figs.\,A1-A5). This 
indicates that Ca\,{\sc ii} and Na\,{\sc i} absorption
arises in the same physical regions. Neutral
sodium has an ionization potential of only $5.1$ eV
(compared to $11.9$ eV for Ca\,{\sc ii}; e.g., 
Morton 2003) and thus
is believed to serve as tracer for cold
neutral gas inside galaxies (e.g., Crawford et al.\,1992). 
However, Na\,{\sc i} absorption is occasionally found also in 
HVCs in the halo of the Milky Way (Richter et al.\,2005; 
Ben Bekhti et al.\,2008) and in the winds of starbursting 
galaxies (Martin et al.\,2005).
We will further investigate the Na\,{\sc i}/Ca\,{\sc ii}
relation in our absorbers in Sect.\,4.1, where 
we also discuss the ionization and dust properties 
in the gas.

%

\begin{figure}[t!]
\centering
\resizebox{1.0\hsize}{!}{\includegraphics{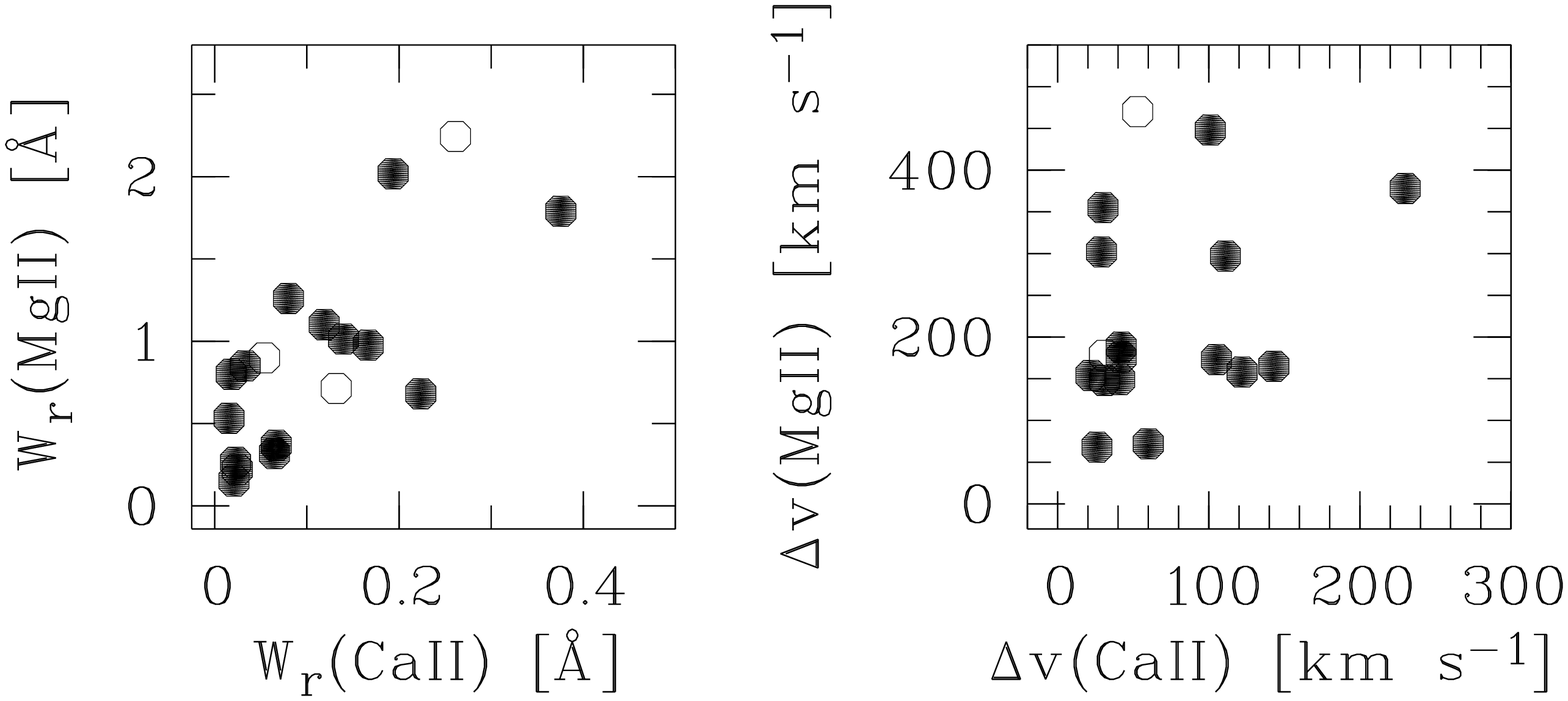}}
\caption[]{Comparison beween Ca\,{\sc ii} and Mg\,{\sc ii}
equivalent and velocity widths in the
intervening Ca\,{\sc ii} absorbers.}
\label{mg2comp}
\end{figure}

%

\subsection{Supplementary information on the absorbers}

%

\begin{table*}[t!]
\caption[]{Supplementary information for selected absorbers in our sample}
\label{dla}
\begin{normalsize}
\begin{tabular}{lccrrrcl}
\hline
QSO & $z_{\rm abs}$ & log $N$(H\,{\sc i})$^{\rm a}$ & Absorber Type & Galaxy Type &
$d$ & $L/L_{\sun}$ & Ref.\\
 & & ($N$ in [cm$^{-2}$]) & & & [kpc] & & \\
 \hline
 J121509+330955 & 0.00396  & 20.34 & DLA     & Early Type &       7.9 & 0.44 & a \\
 J215501-092224 & 0.08091  & 17.98 & LLS     & Spiral     &      34.0 & 0.50 & b \\
 J044117-431343 & 0.10114  & 20.00 & sub-DLA & Disk       &       6.8 & 1.00 & c,d,e \\
 J095456+174331 & 0.23782  & 21.32 & DLA     & Dwarf LSB  & $\leq$3.9 & 0.02 & f\\
 J113007-144927 & 0.31273  & 21.71 & DLA     & Group (tidal debris) & $17-241$ & ... & f,d,g,h\\
 J123200-022404 & 0.39498  & 20.75 & DLA     & Irr. LSB   &    6.6    & 0.17 & i\\
 J045608-215909 & 0.47439  & 19.50 & sub-DLA & ...        & ...       & ...  & j\\
\hline
\end{tabular}
\noindent
\\
$^{\rm a}$\, from H\,{\sc i} Ly\,$\alpha$ absorption\\
References: (a) Miller et al.\,(1999); (b) Jenkins et al.\,(2003); (c) Petitjean et al.\,(1996);
(d) Chen \& Lanzetta (2003); (e) Chen et al.\,(2005); (f) Rao et al.\,(2003); (g) Lane et al.\,(1998);
(h) Kacprzak, Murphy \& Churchill (2010b); (i) Le Brun et al.\,(1997); (j) Turnshek \& Rao (2002)
\end{normalsize}
\end{table*}

%

\begin{figure*}[t!]
\centering
\resizebox{0.9\hsize}{!}{\includegraphics{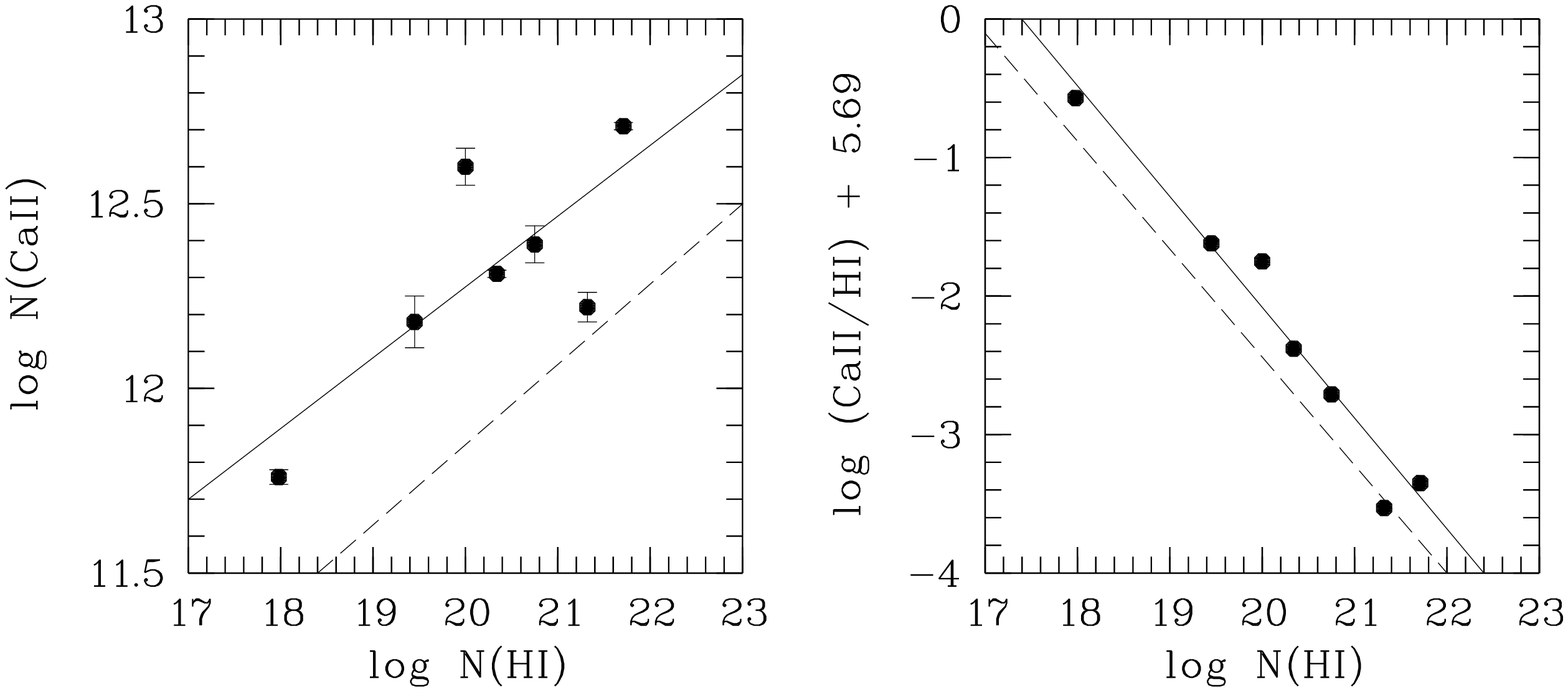}}
\caption[]{{\it Left panel:} relation between Ca\,{\sc ii} and H\,{\sc i} column
densities for seven Ca\,{\sc ii} absorption systems. The solid line indicates
the best fit through the data points, the dashed line indicates the relation
found for Ca\,{\sc ii} and H\,{\sc i} in the Milky Way by Wakker \& Mathis (2000).
{\it Right panel:} relation between log (Ca\,{\sc ii}/H\,{\sc i}), scaled to 
the solar (Ca/H) ratio, and log $N$(H\,{\sc i}) for the same absorbers. The solid 
line again indicates the best fit through the data points, while the dashed line shows
the trend found in the Milky Way (Wakker \& Mathis 2000).}
\end{figure*}

%

For several of our Ca\,{\sc ii} absorbers, supplementary information
on the properties of these systems (based on previous studies) 
is available in the literature. Because intervening
Ca\,{\sc ii} absorption most likely arises in the 
inner and outer regions of galaxies, we are mostly interested in information
that indicates a possible absorber-galaxy connection.
Table 3 lists relevant information for seven out of the 23 QSO sightlines, 
such as the type of the galaxy that probably hosts the 
absorbing cloud in its environment (Col.\,5), the proper absorber-galaxy
distance (``impact parameter''; Col.\,6), and the galaxy luminosity
(Col.\,7). References are given in the last column (see also Zwaan
et al.\,2005). The total neutral gas column density for the 
associated H\,{\sc i} Ly\,$\alpha$ absorption is also listed (Col.\,3), allowing us to 
distinguish between damped Lyman $\alpha$ absorbers (DLAs; log $N$(H\,{\sc i}$)>20.3$),
sub-damped Lyman $\alpha$ absorbers (sub-DLAs; $19.2\leq$\,log $N$(H\,{\sc i}$)\leq20.3$)
and Lyman-limit systems (LLS; $17.2\leq$\,log $N$(H\,{\sc i}$)\leq19.2$),
following the usual H\,{\sc i} Ly\,$\alpha$ absorber classification.
Four out of the seven Ca\,{\sc ii} absorbers are DLAs, two are sub-DLAs, and
there is one LLS. The derived impact parameters are typically small ($d<8$~kpc
for four sightlines out of seven, for which information on $d$ is available).
Note, however, that
neither the H\,{\sc i} column density range nor the distribution of impact
parameters for the absorbers listed in Table 3 are statistically representative 
for intervening Ca\,{\sc ii} absorbers. The data for five out of the seven systems 
are not taken from a random absorber sample, but
come from targeted observations of DLAs and sub-DLAS at low redshift.
Thus, we expect that Table 3 is strongly biased towards high column density 
systems and low impact parameters. Still, the supplementary information listed
in Table 3 provides important information on the physical conditions in these
systems, as will be discussed below.

In Fig.\,9 we show the relation between log $N$(Ca\,{\sc ii}) and 
log $N$(H\,{\sc i}) for these seven systems (left panel), as well
as the relation between the log (Ca\,{\sc ii}/H\,{\sc i}$)=$ 
log ($N$(Ca\,{\sc ii})/$N$(H\,{\sc i})) scaled to the
solar (Ca/H) ratio
and log $N$(H\,{\sc i}) (right panel). Despite the scatter
there is a clear correlation between the Ca\,{\sc ii} column 
density and the H\,{\sc i} column density, which can be fitted as

%

\begin{equation}
{\rm log}\,N({\rm Ca\,II})=8.415+0.193\,[{\rm log}\,N({\rm H\,I})],
\end{equation}

%

in Fig.\,9 (left panel) indicated as solid line. The slope of this
relation agrees remarkably well with the slope found for the
Ca\,{\sc ii}/H\,{\sc i} ratio in Milky Way disk and
halo clouds (Wakker \& Mathis 2000),
but the Milky Way data points are systematically offset 
in either the x or y direction. Most likely, this offset is 
in y and is caused by an above average dust-to-gas ratio in the Milky Way, 
compared to intervening absorbers. This would 
cause a higher depletion of Ca into dust grains, 
so that the measured gas phase Ca column densities 
in the Milky Way absorbers would be systematically lower.
The observed relation indicates that it needs an increase of 
about five orders of magnitude in the total neutral gas 
column density (from $10^{17}$ to $10^{22}$ cm$^{-2}$) 
to lift the gas phase Ca\,{\sc ii} column density by one order of 
magnitude (from $10^{11.7}$ to $10^{12.7}$ cm$^{-2}$). 
As a consequence, the Ca\,{\sc ii}/H\,{\sc i} ratio
decreases strongly with increasing H\,{\sc i} column
density (Fig.\,8, right panel). This tight relation 
in our data can be approximated by the fit

%

\begin{equation}
{\rm log\,(Ca\,II/H\,I)}=8.374-0.807\,[{\rm log}\,N({\rm H\,I})],
\end{equation}

%

as shown as solid line in the right panel of Fig.\,9.
The observed decline
of Ca\,{\sc ii}/H\,{\sc i} over three 
orders of magnitude as a function of $N$(H\,{\sc i})
cannot be primarily caused by a Ca/H abundance
gradient among the absorbers, but is instead
a clear sign of a column-density-dependent 
dust depletion of Ca in the gas. This effect
is well known from observations of Ca\,{\sc ii} and
other ions in the interstellar gas in the Milky Way.
Indeed, the Ca\,{\sc ii}/H\,{\sc i} vs. $N$(H\,{\sc i}) relation 
from Milky Way disk and halo observations (Wakker \& Mathis 2000; 
dashed line) reflects exactly the trend that is seen 
in our intervening Ca\,{\sc ii} absorbers (albeit
the small relative offset; see above).
As we will discuss in Sect.\,4.2, additional evidence 
for dust depletion in the Ca\,{\sc ii} absorbers comes
from the observed Na\,{\sc i}/Ca\,{\sc ii} ratios.

%

\section{Nature and origin of intervening Ca\,{\sc ii} absorbers}

\subsection{Ca\,{\sc ii} absorbers as HVC analogs}

Using Eq.\,(4), together with the observed
Ca\,{\sc ii} column densities (and their uncertainties),
one can calculate that
$3-5$ of the 10 unbiased Ca\,{\sc ii} absorbers would
have H\,{\sc i} column densities in the DLA range
(log $N$(H\,{\sc i}$)\geq 20.3$) , so
that an indirect estimate for the DLA number density
at $z\leq0.5$ in our data is $d{\cal N}/dz$(DLA$)=0.035-0.058$.
This agrees well with direct estimates for
the DLA number density at low redshift ($d{\cal N}/dz$(DLA$)=0.045$;
Zwaan et al.\,2005). With $d{\cal N}/dz=0.117$ (Sect.\,3.2),
Ca\,{\sc ii} absorbers with log $N$(Ca\,{\sc ii}$)\geq 11.65$
outnumber DLAs at low $z$ by a factor of $2-3$. This implies
that Ca\,{\sc ii} absorbers have a higher
absorption cross section than DLAs, i.e., they trace
neutral gas that spans a larger column density range
and that is spatially more extended than DLAs.
Since DLAs at low $z$ are believed to arise in the inner regions of
galaxies (i.e., in their interstellar media), we
conclude that Ca\,{\sc ii} absorbers, which trace gas below
the DLA limit (log $N$(H\,{\sc i}$)<20.3$),
arise predominantly in the outer, more extended regions
of these galaxies, i.e., in their gaseous halos.

Strong observational support for this scenario comes from
the distribution of the H\,{\sc i} 21cm emission in the
Milky Way disk and halo. The 21cm
emission measured in the disk from the position of the
sun indicates a neutral gas column density range that
would cause the disk to appear as a DLA if observed from
outside. In contrast, the majority of the neutral gas clouds in the
Milky Way halo with vertical distances $>1$ kpc
from the Galactic plane (this would include probably
all HVCs and some of the IVCs; see Wakker 2001; Richter 2006)
have H\,{\sc i} column densities log $N$(H\,{\sc i}$)<20.3$
and would therefore be classified as sub-DLAs and LLSs if seen as intervening
absorbers. Moreover, Ca\,{\sc ii} absorption is frequently
detected in the Galactic IVCs and HVCs at column densities
that are very similar to the ones derived by us for the intervening
Ca\,{\sc ii} absorber population (Richter et al.\,2005;
Ben Bekhti et al.\,2008; Richter et al.\,2009; Wakker et al.\,2007,2008).

Based on these arguments, we are led to suggest that
{\it more than half of the Ca\,{\sc ii} absorbers in our
sample trace neutral and partly ionized gas clouds in the halos
and circumgalactic environment of galaxies and thus represent
distant HVC analogs}.

%

\subsection{The Na\,{\sc i}/Ca\,{\sc ii} ratio and dust depletion}

For the eleven Ca\,{\sc ii} absorbers, for which information
on Na\,{\sc i} absorption is available (Sect.\,3.6), we find
a wide range in Na\,{\sc i}/Ca\,{\sc ii} ratios (or upper limits) 
from log (Na\,{\sc i}/Ca\,{\sc ii}$)=-1.36$ 
to log (Na\,{\sc i}/Ca\,{\sc ii}$)=+0.66$,
as listed in Table 4. 
In the Milky Way, these ratios are typical of the diffuse,
warm neutral medium (WNM; $T=10^2-10^4$ K; $n_{\rm H}\leq 10$ cm$^{-3}$), 
where Ca\,{\sc ii} and Na\,{\sc i} often are not the dominant 
ionization states, but serve as trace species 
(see, e.g., Crawford 1992; Welty, Morton \& Hobbs 1996). 
For comparison, the Na\,{\sc i}/Ca\,{\sc ii} ratios that 
are typically found in the Milky Way in the more dense, 
cold neutral medium (CNM; $T<10^2$ K; $n_{\rm H}> 10$ cm$^{-3}$)
lie in the range log (Na\,{\sc i}/Ca\,{\sc ii}$)=1.0-2.5$.
The Na\,{\sc i}/Ca\,{\sc ii} ratios are higher in the CNM than
in the WNM because of the enhanced dust depletion of Ca in the CNM.
It is known that the WNM has a larger volume- and area-filling factor
than the CNM and the molecular gas phase in galaxies and DLAs
(Hirashita et al.\,2003). 
The observed range in the Na\,{\sc i}/Ca\,{\sc ii} ratios 
in the intervening Ca\,{\sc ii} absorbers
thus reflects the large absorption cross section of the WNM
in the inner and outer regions of galaxies
compared with the small cross section of the CNM and
and the molecular gas.

To learn more about the physical properties and origin
of the intervening Ca\,{\sc ii} absorbers we analyse the
implications of their measured Na\,{\sc i}/Ca\,{\sc ii} ratios in 
more detail.
We start by a comparison with the physical conditions in the inner
regions of galaxies (e.g., in gaseous disks), where the gas densities
and dust abundances are expected to be the highest and where
the neutral hydrogen column densities are comparable 
to those in DLAs (i.e., log $N$(H\,{\sc i}$)\geq20.3$).
Under these conditions, the observed Na\,{\sc i}/Ca\,{\sc ii} ratio 
is determined by the gas-phase abundances of Na and Ca and the 
photoionization balance of these ions (i.e., ignoring collisional
ionization). The ratio therefore depends on the {\it local} interstellar 
radiation field, the optical depth in the cloud (often expressed by $A_V$),
the local electron density, and the dust depletion 
of both elements. For the WNM, where the electron densities 
are low ($n_{\rm e}\leq 0.05$ cm$^{-3}$, typically), 
it has been derived from ionization models
that the Na\,{\sc i}/Ca\,{\sc ii}
ratio in dust-free gas is expected to be nearly constant at
log (Na\,{\sc i}/Ca\,{\sc ii}$)\approx -1.6$ (Crawford 1992; 
Welty et al.\,1996; Asplund, Grevesse \& Sauval 2005).
Consequently, the large range in the observed
Na\,{\sc i}/Ca\,{\sc ii} ratios in the WNM in
the Milky Way (log (Na\,{\sc i}/Ca\,{\sc ii}$)\approx -1$ to $+1$;
e.g., Welty et al.\,1996) is caused predominantly by the local 
differences in the Ca dust depletion, but not by ionization 
effects.

Because some of the intervening Ca\,{\sc ii}
absorbers appear to arise in the inner regions of galaxies 
with physical conditions that may be comparable to those in the 
Milky Way disk, we use below the relations discussed above
to provide a rough estimate for the Ca depletion, $\delta_{\rm Ca}$, 
in these systems, where we define
log $\delta_{\rm Ca}=$\,log\,(Ca/H)$_{\rm gas\,phase}-
$\,log (Ca/H)$_{\rm total}$.
If we adopt log (Na\,{\sc i}/Ca\,{\sc ii}$)=-1.6$ for warm neutral 
gas without dust depletion, we can use the measured Na\,{\sc i}/Ca\,{\sc ii}
ratios in the absorbers to estimate $\delta_{\rm Ca,disk}$ in the way
log $\delta_{\rm Ca,disk}=-$\,log\,(Na\,{\sc i}/Ca\,{\sc ii}$)-1.6$,
assuming that Na is not depleted.
From this, we obtain values for log $\delta_{\rm Ca,disk}$ 
(or upper limits) in the range $-0.24$ to $-2.26$, as listed in the
fourth column of Table 4. These values indicate 
a weak to moderate dust depletion of calcium in these absorbers.
The large scatter in $\delta_{\rm Ca}$ most likely reflects
the differences in the local dust abundances in the absorbers
and the large range in physical conditions that are believed to 
balance the formation and destruction of interstellar dust grains.

Note that the values for log $\delta_{\rm Ca}$ listed in Table
4 would be further decreased (by log $\delta_{\rm Na}$), 
if Na is depleted, too.
For the three systems towards J121509+330955, J215501-092224, and
J044117-431343, for which there is information on Ca\,{\sc ii},
Na\,{\sc i}, and H\,{\sc i}, the estimated values for log $\delta_{\rm Ca}$
are generally higher than the values for log (Ca\,{\sc ii}/H\,{\sc i}$)+5.69$,
as shown in Fig.\,9. This is not surprising, however, because
the latter relation depends not only on the dust depletion 
of Ca in the gas, but also on the metallicity of the absorbers,
which are likely to be sub-solar (i.e., log (Ca/H$)<-5.69$).

%

\begin{table}[t!]
\caption[]{Na\,{\sc i}/Ca\,{\sc ii} ratios and Ca depletion}
\label{naca}
\begin{scriptsize}
\begin{tabular}{lcrrr}
\hline
QSO & $z_{\rm abs}$ & log\,$\left( \frac{\rm Na\,I}{\rm Ca\,II} \right) $\,$^{\rm a}$ & log $\delta_{\rm Ca,disk} $\,$^{\rm b}$ & 
log $\delta_{\rm Ca,halo} $\,$^{\rm c}$ \\
\hline
J121509+330955 & 0.00396  & $-0.67$       &  $-0.93$         & $-0.53$ \\ 
J133007-205616 & 0.01831  & $+0.24$       &  $-1.84$         & $-1.44$ \\
J215501-092224 & 0.08091  & $\leq -1.27$  &  $\geq -0.33$    & $0$ \\   
J044117-431343 & 0.10114  & $-0.35$       &  $-1.25$         & $-0.85$ \\ 
J235731-112539 & 0.24763  & $-0.29$       &  $-1.31$         & $-0.91$ \\
J000344-232355 & 0.27051  & $\leq -1.36$  &  $\geq -0.24$    & $0$ \\
J142249-272756 & 0.27563  & $\leq -1.17$\ &  $\geq -0.43$    & $\geq -0.03$ \\
J110325-264515 & 0.35896  & $+0.66$       &  $-2.26$         & $-1.86$ \\
J050112-015914 & 0.40310  & $\leq -1.06$  &  $\geq -0.54$    & $\geq -0.14$ \\
J220743-534633 & 0.43720  & $\leq -0.68$  &  $\geq -0.92$    & $\geq -0.52$ \\
J044117-431343 & 0.44075  & $\leq -0.62$  &  $\geq -0.98$    & $\geq -0.58$\\
\hline
\end{tabular}
\noindent
\\
$^{\rm a}$\,log $\left(\frac{\rm Na\,I}{\rm Ca\,II}\right)=
$log\,$(N($Na\,{\sc i}$)/N($Ca\,{\sc ii}))\\
$^{\rm b}$ Logarithmic depletion of Ca into dust grains, 
assuming Milky Way disk model and zero depletion of Na (see Sect.\,4.2)\\
$^{\rm b}$ Logarithmic depletion of Ca into dust grains, 
assuming Cloudy halo model 
and zero depletion of Na (see Sect.\,4.3)\\
\end{scriptsize}
\end{table}

%

\subsection{Cloudy modelling}

For gas that resides in the inner regions
of galaxies and that has high neutral gas 
column densities, a more
detailed ionization modelling of the ions seen in
absorption is not meaningful because of the mostly
unknown local physical conditions and the
unknown spatial distribution of the gas along the line 
of sight. Since a large
fraction of intervening Ca\,{\sc ii} absorbers are
expected to represent HVC analogs (see Sect.\,4.1),
however, it is useful to investigate 
ion ratios expected for isolated interstellar gas clouds
in the halos and circumgalactic environment of galaxies
that are photoionized by the ambient extragalactic UV
radiation field (see also Richter et al.\,2009).
We therefore modelled the ionization conditions 
of sub-DLAs and LLS using the 
photoionization code Cloudy (v96; Ferland et al.\,1998).
The model absorbers (that are assumed to be optically
thin in H\,{\sc i}) are treated as plane-parallel
slabs, with fixed neutral gas column densities and 
zero dust depletion, which are exposed to the UV 
background radiation. For our study, we considered
a grid of Cloudy models with absorbers 
at $z=0.1,0.3,0.5$ with log $N$(H\,{\sc i}$)=17.0,17.5,18.0,
18.5,19.0,19.5$ and metallicities of $0.1,0.5,1.0$ solar.

For our modelling we focused on the relation between
the column densities of H\,{\sc i}, H\,{\sc ii}, Ca\,{\sc i}, 
Ca\,{\sc ii}, Ca\,{\sc iii}, Na\,{\sc i}, and
Mg\,{\sc ii} as a function of the ionization parameter $U$,
the ratio between the ionizing photon density
and the total particle density (i.e., $U=n_{\gamma}/n_{\rm H}$).
For an assumed ionizing radiation field one can calculate
$n_{\gamma}$ and thus can relate $U$ to the gas density $n_{\rm H}$.
The redshift-dependent UV background was modelled based on
the results by Haardt \& Madau (2001), with solar reference
abundances from Asplund, Grevesse \& Sauval (2005).

In Fig.\,10 we show as an example the Cloudy model for an
absorber at $z=0.3$ with a neutral hydrogen column density of
log $N$(H\,{\sc i}$)=18.0$, and solar calcium and sodium
abundances. In the density range considered
($-2.8 \leq$log\,$n_{\rm H}\leq 0.0$), hydrogen is
predominantly ionized in these model absorbers 
(this is observed also in the Milky Way HVCs; e.g., Tripp et al.\,2003). 
Because we fix $N$(H\,{\sc i}) in our calculations, 
the column density of ionized hydrogen
($N$(H\,{\sc ii})) and the total mass of a model absorber 
increases with decreasing gas density, $n_{\rm H}$, 
because of the increasing hydrogen ionization fraction.
All our Cloudy models show
that both the Ca\,{\sc ii} and the Na\,{\sc i}
column density are nearly constant over the plotted
volume density range and thus both ions follow H\,{\sc i} one-to-one.
In the model shown in Fig.\,10 the 
column density ratios are log (Ca\,{\sc ii}/H\,{\sc i}$)\approx
-5.7$ ($\sim$ (Ca/H$)_{\sun}$) and
log (Na\,{\sc i}/H\,{\sc i}$)\approx -7.0$ ($\sim 0.04\,($Na/H$)_{\sun}$).
The underabundance of Na\,{\sc i} is explained by the fact that
most of the Na is ionized to Na\,{\sc ii}
at these relatively low gas densities. The predicted
Na\,{\sc i}/Ca\,{\sc ii} column density ratio 
is log (Na\,{\sc i}/Ca\,{\sc ii}$)\approx -1.2$, thus
slightly higher than the ratio derived for warm,
dust-free neutral gas clouds (WNM) in the Milky Way disk
(see previous subsection). In analogy to what has been
derived for the Milky Way disk model,
we estimate for the halo model the depletion of Ca using
log $\delta_{\rm Ca,halo}=-$\,log\,(Na\,{\sc i}/Ca\,{\sc ii}$)-1.2$,
assuming that Na is not depleted. 
The derived values for log $\delta_{\rm Ca,halo}$ range
between 0 and $-1.86$ (Table 4, fifth column).

From our Cloudy modelling it follows that Ca\,{\sc ii}
absorbers with log $N$(Ca\,{\sc ii}$)\geq 11.5$
represent reliable tracers for 
neutral and partly ionized gas clouds with
H\,{\sc i} column densities log $N$(H\,{\sc i}$)\geq 17.4$.
However, whether or not Ca\,{\sc ii} is actually detected
in an absorber depends not only on the neutral gas column
density and the Ca abundance, but also on the level of 
dust depletion of Ca. For a dusty multi-phase absorber that has 
both predominantly neutral and predominantly ionized
regions this implies that Ca\,{\sc ii} absorption 
is detectable only in those regions, where the 
metallicity and the neutral gas column density 
is high enough to compensate for the local dust depletion effect.
Compared to Ca, dust depletion is less important for Na,
in particular in diffuse gas (Crawford 1992).
Therefore, the presence or absence of Na\,{\sc i} absorption 
in intervening Ca\,{\sc ii} systems
depends predominantly on the total neutral 
gas column density in the absorbers.
From our Cloudy modelling we find that 
in clouds with sub-solar abundances,
Na\,{\sc i} absorption with log $N$(Na\,{\sc i}$)\geq 11.5$
is detectable only in systems with log $N$(H\,{\sc i}$)>18.5$.
The presence or absence of Na\,{\sc i} absorption in Ca\,{\sc ii} 
absorbers thus discriminates between high-column-density 
H\,{\sc i} clouds (DLAs and sub-DLAs) and low-column-density 
H\,{\sc i} clouds (LLS), respectively.
This explains why only half of the Ca\,{\sc ii}
absorbers show associated Na\,{\sc i} absorption.

As can be seen in Fig.\,10, the Mg\,{\sc ii} column density is, 
in contrast to Ca\,{\sc ii} and Na\,{\sc i}, not constant over 
the plotted density 
range, but increases by about one order of magnitude with decreasing
volume density and increasing total gas column density, 
$N$(H\,{\sc i}+H\,{\sc ii}). Thus, Mg\,{\sc ii} traces both neutral 
and ionized gas. This, together with the
higher cosmic abundance and the less severe dust depletion of Mg,
compared to Ca, makes the Mg\,{\sc ii} ion
more sensitive to the overall gas distribution in the absorbers. 
This explains why the Mg\,{\sc ii} absorption is more complex and
more extended than that of Ca\,{\sc ii} (see Sect.\,3.6).

%

\subsection{Constraints on the gas densities}

From the Milky Way it is known that absorption components with large
Na\,{\sc i}/Ca\,{\sc ii} ratios represent regions 
that have relatively high gas densities. 
As demonstrated above, this effect
is not because of the photoionization balance between
Na\,{\sc i} and Ca\,{\sc ii} in disk and halo clouds, 
but is a result of the density-dependend dust depletion
of Ca. A larger dust depletion of Ca in high-density environments
is expected because of the density-dependent Ca adsorption 
onto grain surfaces (Barlow 1978).
For the Milky Way disk, Crawford (1992) found
that for log (Na\,{\sc i}/Ca\,{\sc ii}$)\sim1$ a typical
gas density is $n_{\rm H}\approx 10$ cm$^{-3}$. 
In the intervening Ca\,{\sc ii} absorbers we always find 
log (Na\,{\sc i}/Ca\,{\sc ii}$)<0.7$ (and in most cases
log (Na\,{\sc i}/Ca\,{\sc ii}$)<0$; see Table 4). As we
have no further direct information on the gas density, 
we regard $n_{\rm H}=10$ cm$^{-3}$ as a realistic upper
limit for the gas densitiy in the Ca\,{\sc ii} absorbers 
detected in our survey.

An estimate for the lower limit of $n_{\rm H}$
in the Ca\,{\sc ii} absorbers can be provided based
on results from ionization modelling of local, low-density
Ca\,{\sc ii} absorbing environments, such as the HVCs in the
Milky Way halo. Ca\,{\sc ii} absorption is frequently observed
in the Galactic H\,{\sc i} 21cm IVCs and HVCs in the Milky Way 
(e.g., West et al. \,1985; D'Odorico et al.\,1989; 
Wakker 2001; Ben Bekhti et al.\,2008,2009). Indeed, this ion plays
a key role in determining the distances of Galactic HVCs and IVCs
(Wakker et al.\,2007,2008; Thom et al.\,2008). Also, below the
detection threshold of H\,{\sc i} 21cm all-sky surveys, Ca\,{\sc ii}
absorption is found in the halo in low column density gas
fragments (Richter et al.\,2005; Ben Bekhti et al.\,2009).
Richter et al.\,(2009) detected Ca\,{\sc ii} absorption, along
with other low ions (e.g., O\,{\sc i} and Si\,{\sc ii}), in two components of
a LLS (=low column density HVC) in the Milky Way halo towards
the QSO PHL\,1811,
at a neutral gas column density of only log $N($H\,{\sc i}$)\approx 17$.
From detailed ionization modelling,
a lower limit for the density of $n_{\rm H}\geq0.3$ cm$^{-3}$
was derived for one of those absorption components, and a limit of
$n_{\rm H}\geq0.05$ cm$^{-3}$ was obtained for the other
component. A value of $n_{\rm H}\approx 0.05$ cm$^{-3}$ 
probably also represents a realistic lower limit for the gas 
density in intervening Ca\,{\sc ii} absorbers.

%

\begin{figure}[t!]
\resizebox{1.0\hsize}{!}{\includegraphics{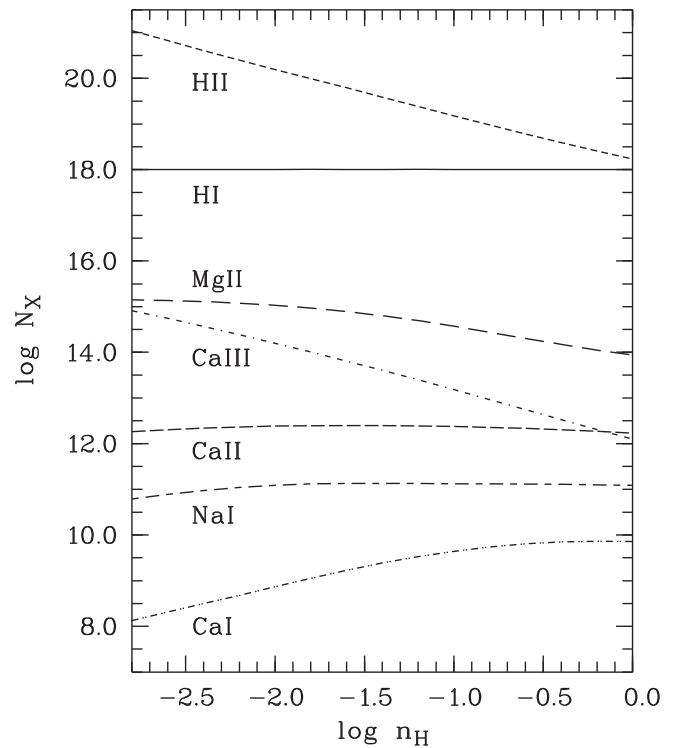}}
\caption[]{
Cloudy model for an example of an absorber at $z=0.3$
with a neutral hydrogen column density of
log $N$(H\,{\sc i}$)=18.0$ and solar calcium and sodium
abundances. Shown are the expected ion column densities
of the ions H\,{\sc i}, H\,{\sc ii}, Ca\,{\sc ii}, Mg\,{\sc ii}, and
Na\,{\sc i} as a function of the gas density.
}
\end{figure}

%

\subsection{The size of neutral gas halos around galaxies}

Assuming that intervening Ca\,{\sc ii} absorption arises in both the
disks and extended gaseous halos of galaxies, 
we can use the relation 

%

\begin{equation}
\frac{d{\cal N}}{dz}=
\frac{n_{\rm g}\,\langle
f_{\rm c} \rangle \,c\,\pi R_{\rm h}^2}
{H(z)}
\end{equation}

%

between the geometrical cross 
section of the absorbers and the absorber number density 
per unit redshift to investigate the spatial extent
of galaxies and their gaseous halos (e.g., Kacprzak et al.\,2008). 
In the above equation, $n_{\rm g}$ is the space density of galaxies,
$R_{\rm h}$ is the radius of the Ca\,{\sc ii} absorbing region
in a galaxy (i.e., disk+halo), $f_{\rm c,Ca\,II}\leq 1$ is the covering 
fraction of Ca\,{\sc ii} absorbing gas above a given column density limit, and
$H(z)=H_0\,(\Omega_{\rm m}\,(1+z)^3+\Omega_{\Lambda})^{1/2}$ is
the Hubble parameter.
The space density of galaxies above a given luminosity 
threshold $L_{\rm min}$ can be obtained from the Schechter
galaxy luminosity function. For our further considerations
we assume $n_{\rm g}=0.0107$\,Mpc$^{-3}$ for 
$L_{\rm min}=0.05\,L^{\star}$ and $\langle z \rangle \approx 0.35$,
based on the results of Faber et al.\,(2007).

Unfortunately, the covering fraction of Ca\,{\sc ii} in galaxy halos 
is not well constrained. While Ben Bekhti et al.\,(2008) have found
$f_{\rm c,Ca\,II}\approx 0.22$ for log $N$(Ca\,{\sc ii}$)\geq 11.65$ 
for the Milky Way halo based on the same
UVES optical data that we use in this paper, this result cannot be 
directly adopted here because of the different observing perspectives 
(inside view vs.\,outside view).
Modelling the radial distribution of Ca\,{\sc ii} absorbers
in galaxy halos based on all available observational results clearly is
beyond the scope of this study. This issue will be addressed in detail
in a subsequent paper, but here we present some simple considerations about the
absorption cross section of Ca\,{\sc ii} in galaxies.

First, let us assume that all galaxies have neutral gas disks with
radius $R_{\rm d}$ that represent DLAs and that the disks
are spherically surrounded by a population of neutral gas 
clouds (sub-DLAs and LLS) in a halo with radius $R_{\rm h}$. 
Second, we assume the simplest case of gaseous disks 
with a Ca\,{\sc ii} covering fraction of $f_{\rm c,Ca\,II}=1$, and with a total 
absorption cross
section of $\sigma_{\rm d}=\pi R_{\rm d}^2$ when seen face-on and 
$\sigma_{\rm d}=0$ when seen edge-on. For the halo cloud population we assume that
the {\it volume} filling factor of neutral gas in the halo is constant, so
that the observed covering fraction scales directly with the
absorption pathlength through the halo. Moreover, we fix the Ca\,{\sc ii} 
covering fraction for log $N$(Ca\,{\sc ii}$)\geq 11.65$
along the pathlength $R_{\rm h}$ (inside view) to $0.22$, based on the results
of Ben Bekhti et al.\,(2008) for the Milky Way halo. Because the mean pathlength
through a spherical halo from outside is $\sim 1.23$ times the radius
of that sphere, we assume that the mean Ca\,{\sc ii} covering fraction
in a galaxy halo
(for log $N$(Ca\,{\sc ii}$)\geq 11.65$) from an external vantage point is
$f_{\rm c,Ca\,II}=1.23\times 0.22\approx 0.27$. The total absorption cross 
section of the neutral halo gas then is 
$\sigma_{\rm h}=\left< f_{\rm c,Ca\,II} \right>\,
\pi R_{\rm h}^2$. Note that we neglect any luminosity scaling for the 
galaxy sizes at this point. The next step in our galaxy halo 
model is to invert Eq.\,(5) and insert the above
value for $n_{\rm g}$, together with $H(z=0.35)$, 
$d{\cal N}/dz$(Ca\,{\sc ii}$)=0.117$,
and $d{\cal N}/dz$(DLA$)=0.045$ and iteratively 
solve for $R_{\rm h}$, $R_{\rm d}$, and 
$\left< f_{\rm c,Ca\,II} \right>$. 
If we do so, we obtain $R_{\rm h}=55$ kpc, 
$R_{\rm d}=20$ kpc and $\left< f_{\rm c,Ca\,II} \right> = 0.33$. 
Thus, if all galaxies with $L\geq0.05\,L^{\star}$ are spherically 
surrounded by a population of Ca\,{\sc ii} absorbing neutral/partly ionized 
gas clouds (i.e., HVCs), e.g., as part of their accretion and/or star-formation
activities, and if the mean Ca\,{\sc ii} covering fraction of this gas 
is $\sim 0.33$, then {\it the 
characteristic radial extent of such a HVC population
is $R_{\rm HVC}\approx 55$ kpc}. 

For comparison, the most distant known HVC in the halo of 
the Milky Way is the Magellanic Stream (MS), which has a distance
of $\sim 50$~kpc from the Galaxy (Gardiner \& Nogudi 1996). The MS 
is a tidal stream that is being 
accreted by the Milky Way (e.g., Fox et al.\,2010). 
For M31, Thilker et al.\,(2004) have shown that the population of 
H\,{\sc i} 21cm halo-clouds around Andromeda extends to radii of $\sim 60$ kpc.
For Mg\,{\sc ii}, Kacprzak et al.\,(2008) estimated a halo size of 
$R_{\rm h}\sim 91$ kpc for $L\geq0.05\,L^{\star}$ galaxies at $z=0.5$, assuming 
$f_{\rm c,Mg\,II}\sim 0.5$ and no luminosity scaling of $R_{\rm h}$. The
larger halo radius for Mg\,{\sc ii} (and the higher covering fraction) 
reflects the higher sensitivity
of Mg\,{\sc ii} absorption for low column density neutral and ionized gas
in the outskirts of galaxies.

It is important to note that
our approach to constrain the sizes of neutral gas
halos of galaxies in the local Universe from the number density
of intervening Ca\,{\sc ii} absorbers is based on very simple 
assumptions. In particular, the covering fraction of neutral
gas in the halos of galaxies may not be constant, but may decrease
with increasing radius (as seen in M31; Thilker et al.\,2004).
Nonetheless, our simplified estimate for $R_{\rm HVC}$ delivers a very plausible
result, which is consistent with what we
know about the distribution of neutral gas around galaxies 
in the local Universe (e.g., Sancisi et al.\,2008).

%

\subsection{Intervening Ca\,{\sc ii} and the H\,{\sc i} CDDF}

In Sect.\,3.4 we investigated the relation between Ca\,{\sc ii}
and H\,{\sc i} in individual systems. We will now 
discuss the frequency of Ca\,{\sc ii} systems in a more 
cosmological context, i.e., by relating the Ca\,{\sc ii}
number density with the CDDF of intervening H\,{\sc i}
absorbers. Because the H\,{\sc i} CDDF for $16\leq$ log 
$N$(H\,{\sc i}$)\leq 21$ is not well constrained for 
$z\leq 0.5$ (see Lehner et al.\,2007 and references therein),
we adopt a ``standard'' CDDF for intervening
H\,{\sc i} absorbers with with $\beta=1.5$, so that 
$f(N)\propto N^{-1.5}$ 
(e.g., Kim et al.\,2002). Let us first consider that every
intervening Ca\,{\sc ii} absorber traces a corresponding H\,{\sc i}
system. For DLAs with log $N$(H\,{\sc i}$)\geq 20.3$
one finds $d{\cal N_{\rm DLA}}/dz=0.045$ at $z\leq0.5$ (e.g., Zwaan 
et al.\,2005). An absorber number density of $d{\cal N}/dz=0.117$
instead corresponds to all absorbers with 
log $N$(H\,{\sc i}$)\geq 19.7$ in the H\,{\sc i} CDDF 
(as derived from integrating the 
H\,{\sc i} CDDF over the appropriate range). From 
Eq.\,(4) and Fig.\,9 we find that the mean
(Ca\,{\sc ii}/H\,{\sc i}) ratio is $\sim 2.4$ dex
below the solar value for absorbers in the range 
$19.7\leq$\,log $N$(H\,{\sc i}$)\leq20.5$ 
(i.e., those absorbers that dominate the number density,
$d{\cal N}/dz$(Ca\,{\sc ii})). This indicates, along with the
solar Ca abundance of (Ca/H$)_{\sun}=-5.69$,
that H\,{\sc i} absorbers with log 
$N$(H\,{\sc i}$)\geq19.7$ should have (on average)
Ca\,{\sc ii} column densities of log $N$(Ca\,{\sc ii}$)\geq 11.6$.

This expectation value excellently agrees with our
completeness limit of log $N$(Ca\,{\sc ii}$)=11.65$, 
for which $d{\cal N}/dz$(Ca\,{\sc ii}) 
was derived. This shows that - if one takes 
dust depletion of Ca into account - the observed
Ca\,{\sc ii} number density of $d{\cal N}/dz=0.117$
for log $N$(Ca\,{\sc ii}$)>11.65$ agrees very well with
with the number of metal enriched sub-DLAs and DLAs
expected from the H\,{\sc i} CDDF. Note that this
correspondence can be interpreted only in a statistical sense.
There are individual systems with log $N$(H\,{\sc i}$)\geq19.7$
and log $N$(Ca\,{\sc ii}$)<11.6$ as well as systems
with log $N$(H\,{\sc i}$)<19.7$ and log $N$(Ca\,{\sc ii}$)\geq 11.6$
(see Tables 2 and 3).

%

\section{Comparison with previous Ca\,{\sc ii} studies}

While Ca\,{\sc ii} was among the first ions detected in 
intervening absorbers in QSO spectra (see Blades 1988 for a review), 
there are only a few studies that have used this ion
in a systematic manner to investigate QSO absorption line
systems. This is because strong,
intervening Ca\,{\sc ii} absorption is apparently
very rare and the detection of the more frequent weak 
Ca\,{\sc ii} absorbers requires high S/N and high
spectral resolution.
Bowen et al.\,(1991) have studied intervening
Ca\,{\sc ii} absorption in the halos of nine low-redshift ($z<0.2$) 
galaxies based on intermediate resolution optical spectra obtained 
with various different instruments. The $2\sigma$ equivalent width
detection limits in the $\lambda 3934$ line of those data
range between $38$ and $164$ m\AA. While all nine sightlines
pass within $45\,h^{-1}$ kpc of the galaxies, only one
detection is reported. Including previous results on 
intervening Ca\,{\sc ii} absorption along QSO sightlines, 
Bowen et al.\,concluded that Ca\,{\sc ii} absorption in 
galaxy halos is relatively weak and the distribution of Ca\,{\sc ii}
around galaxies is inhomogeneous with covering fractions
much smaller than unity for the equivalent width limit reached 
in their survey.
Our results confirm this scenario. From our data we estimate 
that the covering fraction of Ca\,{\sc ii} absorption in the 
halos of galaxies with $W_{\rm r}>70$ m\AA\, in the 
$\lambda 3934$ line, is $\sim 1/9$,
which agrees very well with the detection rate
reported by Bowen et al.\,(1991).

The more recent systematic studies of intervening Ca\,{\sc ii} 
absorption at low and intermediate redshift by Wild, Hewett \& 
Pettini (2006, 2007), Zych et al.\,(2007, 2009), and Nestor
et al.\,(2008) have concentrated on the frequency and nature
of {\it strong} Ca\,{\sc ii} absorption systems with $W_{\rm r,3934}>200$ 
m\AA, motivated by the detection of these systems 
in the very large sample ($\sim 14,500$) of low resolution
spectra from the Sloan Digital Sky Survey (SDSS). From these
studies it was concluded that strong Ca\,{\sc ii} absorbers
represent the subset of DLAs with a particularly high dust content.
From the SDDS data, Wild, Hewett \& Pettini (2006) 
have derived a number density of $d{\cal N}/dz\approx 0.02$ 
for strong Ca\,{\sc ii} systems with $W_{\rm r,3934}\geq 500$ m\AA\, 
at $\left< z_{\rm abs} \right>=0.95$, which is $\sim 20-30$ 
percent of the number density of DLAs at that redshift. In our 
Ca\,{\sc ii} absorber sample there is only one system (in the
unbiased sample) that has $W_{\rm r,3934}\geq 500$ m\AA\,
(see Table 2). Because all 304 spectra in our original 
QSO sample have a S/N high enough
to detect Ca\,{\sc ii} absorption at this level (see Sect.\,2.1),
the total redshift path available for detecting strong Ca\,{\sc ii} absorption in
our data is $\Delta z=100.60$. Thus, $d{\cal N}/dz\approx 0.01$
for $W_{\rm r,3934}\geq 500$ m\AA\, and $\left< z_{\rm abs} \right>=0.35$. 
This is $\sim 20$ percent of the number density of DLAs at low redshift
(Zwaan et al.\,2005), and thus similar to the results for 
$\left< z_{\rm abs} \right>=0.95$ by Wild, Hewett \& Pettini (2006).
Therefore, the relative cross section of strong Ca\,{\sc ii}
systems compared with DLAs has not changed from $z=1$ to $z=0.3$.

%

\section{Summary}

We presented a systematic study of intervening
Ca\,{\sc ii} absorption in the redshift range $z=0.0-0.5$, 
along 304 QSO sightlines with a total redshift path of
$\Delta z\approx 100$, using optical high-resolution spectra obtained
with VLT/UVES. The main results can be summarized as follows:\\
\\
(1) We detect 23 intervening Ca\,{\sc ii} absorbers 
at redshifts $z=0.00396-0.47439$ in our data,
at rest frame equivalent widths $W_{{\rm r},3934}=15-799$
m\AA\, and column densities log $N$(Ca\,{\sc ii}$)=11.25-13.04$.
We determine a bias-corrected number density of Ca\,{\sc ii} 
absorbers per unit redshift of $d{\cal N}/dz$(Ca\,{\sc ii}$)=0.117\pm0.044$ 
for absorbers with log $N$(Ca\,{\sc ii}$)>11.65$. Ca\,{\sc ii} 
absorbers above this column density level 
outnumber damped Lyman $\alpha$ absorbers (DLAs) at low redshift
by a factor of two to three.\\
\\
(2) From ionization modelling we conclude
that intervening Ca\,{\sc ii} absorbers, with 
log $N$(Ca\,{\sc ii}$)>11.5$, trace neutral and partly 
ionized gas at H\,{\sc i} column densities
log $N$(H\,{\sc i}$)>17.4$. Probably more than half
of the detected Ca\,{\sc ii} absorbers arise
in gas clouds at H\,{\sc i} column densities 
below the DLA limit (log $N$(H\,{\sc i}$)\leq20.3)$.
These absorbers therefore mimic the properties 
of the H\,{\sc i} 21cm high-velocity clouds (HVCs) in the 
halo of the Milky Way. This suggests, along with the large absorption
cross section of Ca\,{\sc ii} absorbers compared
with DLAs, that most of the weak Ca\,{\sc ii} systems
represent HVC analogs in the halos of intervening galaxies.\\
\\
(3) The subcomponent structure of intervening Ca\,{\sc ii} absorbers
is relatively simple, typically with $\leq 3$ absorption components
per system. Ca\,{\sc ii} absorption components follow a well defined
column density distribution function (CDDF), 
$f(N)=C\,N^{-\beta}$, with $\beta=1.68\pm0.20$. This CDDF is
mildly steeper than what is typically found for other ions (e.g., Mg\,{\sc ii}),
indicating the effect of enhanced depletion of Ca into dust grains
in high column density absorbers.\\
\\
(4) Almost all (18/19) Ca\,{\sc ii} systems, for which information
on Mg\,{\sc ii} is available, show associated strong Mg\,{\sc ii} absorption 
with Mg\,{\sc ii} rest frame equivalent widths $W_{\rm r,2796}\geq 300$ m\AA.
While the Ca\,{\sc ii} absorption generally is aligned with
the strongest Mg\,{\sc ii} absorption component, the velocity
extent of the Mg\,{\sc ii} absorption is substantially larger
than that of Ca\,{\sc ii}.
We conclude that the absorbers represent extended, multicomponent
and multiphase gaseous structures, for which the Ca\,{\sc ii} absorption 
is detectable only in the regions with the highest neutral gas column 
densities.\\
\\
(5) For seven of the Ca\,{\sc ii} absorbers we have supplementary 
information on the H\,{\sc i} column density from previous studies.
The calcium-to-hydrogen ratio (Ca\,{\sc ii}/H\,{\sc i}) decreases strongly 
with increasing H\,{\sc i} column density (from log 
(Ca\,{\sc ii}/H\,{\sc i}$)=-6.3$ at log $N$(H\,{\sc i}$)=18$
to log (Ca\,{\sc ii}/H\,{\sc i}$)=-9.3$ at log $N$(H\,{\sc i}$)=22$). 
Five Ca\,{\sc ii} absorbers are detected in Na\,{\sc i} (detection
rate is $45$ percent) and the
observed Na\,{\sc i}/Ca\,{\sc ii} ratios lie in the range $0.2-4.6$.
These numbers indicate moreover that Ca\,{\sc ii} absorbers contain substantial 
amounts of dust and that the depletion of Ca into dust grains 
increases with increasing gas column density. This trend mimics
the trend seen in the Milky Way disk and halo gas.\\
\\
(6) Taking into account the observed Ca\,{\sc ii}-H\,{\sc i} relation,
the number density of Ca\,{\sc ii} absorbers of $d{\cal N}/dz=0.117$,
for log $N$(Ca\,{\sc ii}$)>11.65$, matches the expected value for metal- and
dust-enriched sub-DLAs and DLAs estimated from the H\,{\sc i} CDDF.
These results support our original idea that
intervening Ca\,{\sc ii} absorbers represent excellent tracers of
metal-enriched neutral gas in the inner and outer regions of galaxies.\\
\\
(7) From the observed number density of Ca\,{\sc ii}
absorbers per unit redshift of $d{\cal N}/dz$(Ca\,{\sc ii}$)=0.117$,
we estimate the characteristic radial extent, $R_{\rm HVC}$, 
of (partly) neutral gas clouds (HVCs) around low-redshift galaxies
with log $N$(H\,{\sc i}$)\geq 17.4$.
Considering all galaxies with $L\geq0.05L^{\star}$, and
assuming that the mean (projected) Ca\,{\sc ii} covering fraction
inside and outside of galaxies is $\left< f_{\rm c, CaII} \right>=0.33$
(as estimated from Milky Way Ca\,{\sc ii} absorption statistics),
we derive $R_{\rm HVC}\approx 55$ kpc.\\
\\
The study presented in this paper demonstrates that (weak)
intervening Ca\,{\sc ii} absorbers provide important information
on the neutral gas extent of galaxies.
More detailed studies of this absorber population 
thus might be useful to investigate the circulation
processes of neutral gas in galaxy halos that 
characterize the ongoing formation
and evolution of galaxies at low redshift due to galaxy merging,
gas accretion from the IGM, and outflows.

For a better understanding of these processes and to quantify 
the gas accretion rate of galaxies in the local Universe it will be
particularly important to investigate in detail the {\it radial} 
distribution of neutral and ionized gas around galaxies. This can be 
done by combining absorption-line studies in the 
optical and in the UV with high-resolution H\,{\sc i} 21cm observations of
nearby galaxies. Indeed, with the advent of poweful optical multiaperture
spectrographs at ground based 8m-class telescopes, a new
sensitive UV spectrograph installed on HST ({\it Cosmic Origins
Spectrograph}; COS), and new sensitive H\,{\sc i} 21cm surveys 
(e.g., the Effelsberg-Bonn-H\,{\sc i}-Survey; EBHIS), future multiwavelength
studies are expected to provide important new details
on the complex interplay between galaxies and their circumgalactic gaseous 
environment.

%

\begin{acknowledgements}

P.R. acknowledges financial support by the German
\emph{Deut\-sche For\-schungs\-ge\-mein\-schaft}, DFG,
through grant Ri 1124/5-1. J.C.C. is grateful for support
from NASA under grant NAG5-6399 NNG04GE73G.

\end{acknowledgements}

%

%

\newpage

\appendix

\section{Supplementrary tables and figures}

%

\begin{table}[h!]
\caption[]{Results from Voigt-profile fitting of intervening Ca\,{\sc ii}
absorbers, part 1}
\label{fitting1}
\begin{footnotesize}
\begin{tabular}{rrcr}
\hline
Comp. & $v_{\rm rel}$  & log $N$(Ca\,{\sc ii}) & $b$ \\
      & [km\,s$^{-1}$] & ($N$ in [cm$^{-2}$])  & [km\,s$^{-1}$] \\
\hline
\multicolumn{4}{c}{J121509+330955 - $z=0.00396$}\\
\hline
1     & $-73$            & $11.85\pm0.02$      & $19.7\pm3.1$ \\
2     & $-27$            & $11.14\pm0.06$      &  $4.0\pm1.2$ \\
3     & $-2$             & $12.08\pm0.02$      & $11.1\pm0.4$ \\
\hline
\multicolumn{4}{c}{J133007-205616 - $z=0.01831$}\\
\hline
1     & $-303$           & $11.43\pm0.22$      &  $4.0\pm1.6$ \\
2     & $-284$           & $11.73\pm0.07$      & $11.4\pm2.8$ \\
3     & $-233$           & $11.95\pm0.03$      & $10.1\pm1.1$ \\
4     & $-92$            & $12.18\pm0.03$      & $18.7\pm3.2$ \\
5     & $-58$            & $12.48\pm0.02$      & $16.4\pm0.7$ \\
6     & $-8$             & $12.13\pm0.03$      & $15.3\pm1.3$ \\
7     & $+3$             & $11.96\pm0.07$      &  $3.0\pm0.6$ \\
8     & $+20$            & $12.16\pm0.03$      & $12.7\pm1.0$ \\
9     & $+46$            & $12.02\pm0.03$      & $13.3\pm1.2$ \\
\hline
\multicolumn{4}{c}{J215501-092224 - $z=0.08091$}\\
\hline
1     & $+1$             & $11.76\pm0.02$      &  $7.1\pm0.6$ \\
\hline
\multicolumn{4}{c}{J044117-431343 - $z=0.10114$}\\
\hline
1     & $-83$            & $11.54\pm0.03$      &  $3.0\pm0.4$ \\
2     & $-65$            & $11.76\pm0.20$      &  $7.7\pm2.0$ \\
3     & $-55$            & $11.57\pm0.30$      &  $3.1\pm3.0$ \\
4     & $-38$            & $11.30\pm0.08$      &  $5.5\pm1.7$ \\
5     & $-13$            & $11.72\pm0.08$      & $14.7\pm2.4$ \\
6     & $0$              & $11.94\pm0.05$      &  $5.1\pm0.6$ \\
7     & $+14$            & $12.04\pm0.05$      & $16.1\pm1.5$ \\
\hline
\multicolumn{4}{c}{J095456+174331 - $z=0.23782$}\\
\hline
1     & $-22$            & $11.47\pm0.09$      & $16.4\pm3.8$ \\
2     & $-2$             & $11.90\pm0.08$      &  $5.9\pm1.4$ \\
3     & $+13$            & $11.55\pm0.16$      & $21.2\pm7.1$ \\
4     & $+41$            & $11.29\pm0.14$      &  $6.9\pm2.1$ \\
\hline
\multicolumn{4}{c}{J012517-001828 - $z=0.23864$}\\
\hline
1     & $+2$             & $11.48\pm0.06$      &  $3.0\pm0.7$ \\
\hline
\multicolumn{4}{c}{J235731-112539 - $z=0.24763$}\\
\hline
1     & $-13$            & $11.94\pm0.06$      &  $3.3\pm2.2$ \\
2     & $+1$             & $12.49\pm0.03$      &  $7.0\pm1.2$ \\
3     & $+21$            & $12.03\pm0.05$      &  $6.9\pm1.4$ \\
\hline
\multicolumn{4}{c}{J000344-232355 - $z=0.27051$}\\
\hline
1     & $-18$            & $10.88\pm0.06$      &  $8.2\pm1.6$ \\
2     & $+1$             & $11.58\pm0.02$      &  $3.0\pm0.6$ \\
\hline
\multicolumn{4}{c}{J142249-272756 - $z=0.27563$}\\
\hline
1     & $-2$             & $12.07\pm0.03$      &  $6.0\pm0.7$ \\
\hline
\multicolumn{4}{c}{J042707-130253 - $z=0.28929$}\\
\hline
1     & $-105$           & $11.82\pm0.03$      &  $5.1\pm0.6$ \\
2     & $0$              & $11.61\pm0.04$      &  $6.7\pm0.8$ \\
\hline
\multicolumn{4}{c}{J113007-144927 - $z=0.31273$}\\
\hline
1     & $-52$            & $11.83\pm0.02$      &  $9.7\pm0.3$ \\
2     & $-26$            & $11.46\pm0.02$      &  $8.4\pm0.6$ \\
3     & $-16$            & $11.80\pm0.02$      &  $3.1\pm0.4$ \\
4     & $-1$             & $11.92\pm0.02$      &  $3.7\pm0.3$ \\
5     & $+11$            & $11.71\pm0.02$      &  $4.6\pm0.3$ \\
6     & $+30$            & $12.06\pm0.02$      & $10.7\pm0.3$ \\
7     & $+56$            & $11.73\pm0.09$      &  $5.9\pm1.8$ \\
8     & $+62$            & $11.71\pm0.05$      &  $3.0\pm0.8$ \\
\hline
\end{tabular}
\noindent
\\
\end{footnotesize}
\end{table}

\newpage

%
 
\begin{table}[h!]
\caption[]{Results from Voigt-profile fitting of intervening Ca\,{\sc ii}
absorbers, part 2}
\label{fitting2}
\begin{footnotesize}
\begin{tabular}{rrcr}
\hline
Comp. & $v_{\rm rel}$  & log $N$(Ca\,{\sc ii}) & $b$ \\
      & [km\,s$^{-1}$] & ($N$ in [cm$^{-2}$])  & [km\,s$^{-1}$] \\
\hline
\multicolumn{4}{c}{J102837-010027 - $z=0.32427$}\\
\hline
1     & $0$              & $12.38\pm0.02$      & $12.5\pm0.8$ \\
2     & $+19$            & $11.51\pm0.10$      &  $3.0\pm0.3$ \\
\hline
\multicolumn{4}{c}{J231359-370446 - $z=0.33980$}\\
\hline
1     & $-9$             & $12.06\pm0.04$      &  $2.1\pm0.3$ \\
2     & $0$              & $12.58\pm0.05$      &  $2.8\pm0.9$ \\
3     & $+18$            & $11.16\pm0.07$      &  $2.2\pm0.7$ \\
\hline
\multicolumn{4}{c}{J110325-264515 - $z=0.35896$}\\
\hline
1     & $0$              & $11.26\pm0.02$      &  $6.1\pm0.4$ \\
\hline
\multicolumn{4}{c}{J094253-110426 - $z=0.39098$}\\
\hline
1     & $-3$             & $11.71\pm0.02$      &  $3.7\pm0.4$ \\
2     & $+3$             & $11.40\pm0.04$      &  $6.2\pm1.1$ \\
\hline
\multicolumn{4}{c}{J121140-103002 - $z=0.39293$}\\
\hline
1     & $-3$             & $11.38\pm0.06$      &  $9.9\pm1.7$ \\
\hline
\multicolumn{4}{c}{J123200-022404 - $z=0.39498$}\\
\hline
1     & $-20$            & $11.81\pm0.11$      & $14.1\pm2.5$ \\
2     & $-3$             & $11.79\pm0.12$      &  $9.2\pm1.2$ \\
3     & $+22$            & $11.94\pm0.05$      & $19.8\pm2.5$ \\
4     & $+38$            & $11.36\pm0.05$      &  $3.0\pm0.4$ \\
5     & $+52$            & $11.04\pm0.09$      &  $3.3\pm0.7$ \\
\hline
\multicolumn{4}{c}{J050112-015914 - $z=0.40310$}\\
\hline
1     & $0$              & $11.85\pm0.04$      &  $3.0\pm0.3$ \\
2     & $+36$            & $11.89\pm0.05$      &  $5.9\pm1.1$ \\
3     & $+62$            & $11.53\pm0.08$      &  $3.3\pm0.4$ \\
\hline
\multicolumn{4}{c}{J224752-123719 - $z=0.40968$}\\
\hline
1     & $-44$            & $11.61\pm0.05$      &  $4.3\pm1.1$ \\
2     & $-2$             & $12.11\pm0.02$      & $15.6\pm1.3$ \\
\hline
\multicolumn{4}{c}{J220743-534633 - $z=0.43720$}\\
\hline
1     & $-2$             & $11.98\pm0.05$      &  $2.4\pm0.7$ \\
\hline
\multicolumn{4}{c}{J044117-431343 - $z=0.44075$}\\
\hline
1     & $-7$             & $11.16\pm0.04$      &  $3.1\pm0.6$ \\
2     & $+4$             & $10.95\pm0.21$      &  $3.0\pm1.1$ \\
3     & $+28$            & $10.42\pm0.21$      &  $3.3\pm0.8$ \\
\hline
\multicolumn{4}{c}{J144653-011356 - $z=0.44402$}\\
\hline
1     & $0$             & $11.25\pm0.06$      &  $3.2\pm0.4$ \\
\hline
\multicolumn{4}{c}{J045608-215909 - $z=0.47439$}\\
\hline
1     & $-5$             & $12.02\pm0.07$      & $15.8\pm1.4$ \\
2     & $+1$             & $10.72\pm0.46$      &  $2.3\pm0.8$ \\
3     & $+22$            & $11.45\pm0.19$      & $11.6\pm3.8$ \\
4     & $+52$            & $11.07\pm0.13$      &  $6.9\pm3.9$ \\
\hline
\end{tabular}
\noindent
\\
\end{footnotesize}
\end{table}

%

\begin{table}[t!]
\caption[]{List of additional Ca\,{\sc ii} candidate systems}
\label{candidate}
\begin{tabular}{lr}
\hline
QSO & $z_{\rm abs}$ \\
\hline
J162439+234512 & 0.31759 \\
J104733+052454 & 0.31822 \\
J005102-252846 & 0.34393 \\
J142253-000149 & 0.34468 \\
J232820+002238 & 0.41277 \\
J051707-441055 & 0.42913 \\
J055158-211949 & 0.43982 \\
J030844-295702 & 0.44100 \\
J000344-232355 & 0.45241 \\
J124646-254749 & 0.49282 \\
J014125-092843 & 0.50053 \\
\hline
\end{tabular}
\noindent
\\
\end{table}

%

\newpage

\begin{figure*}[t!]
\centering
\resizebox{0.9\hsize}{!}{\includegraphics{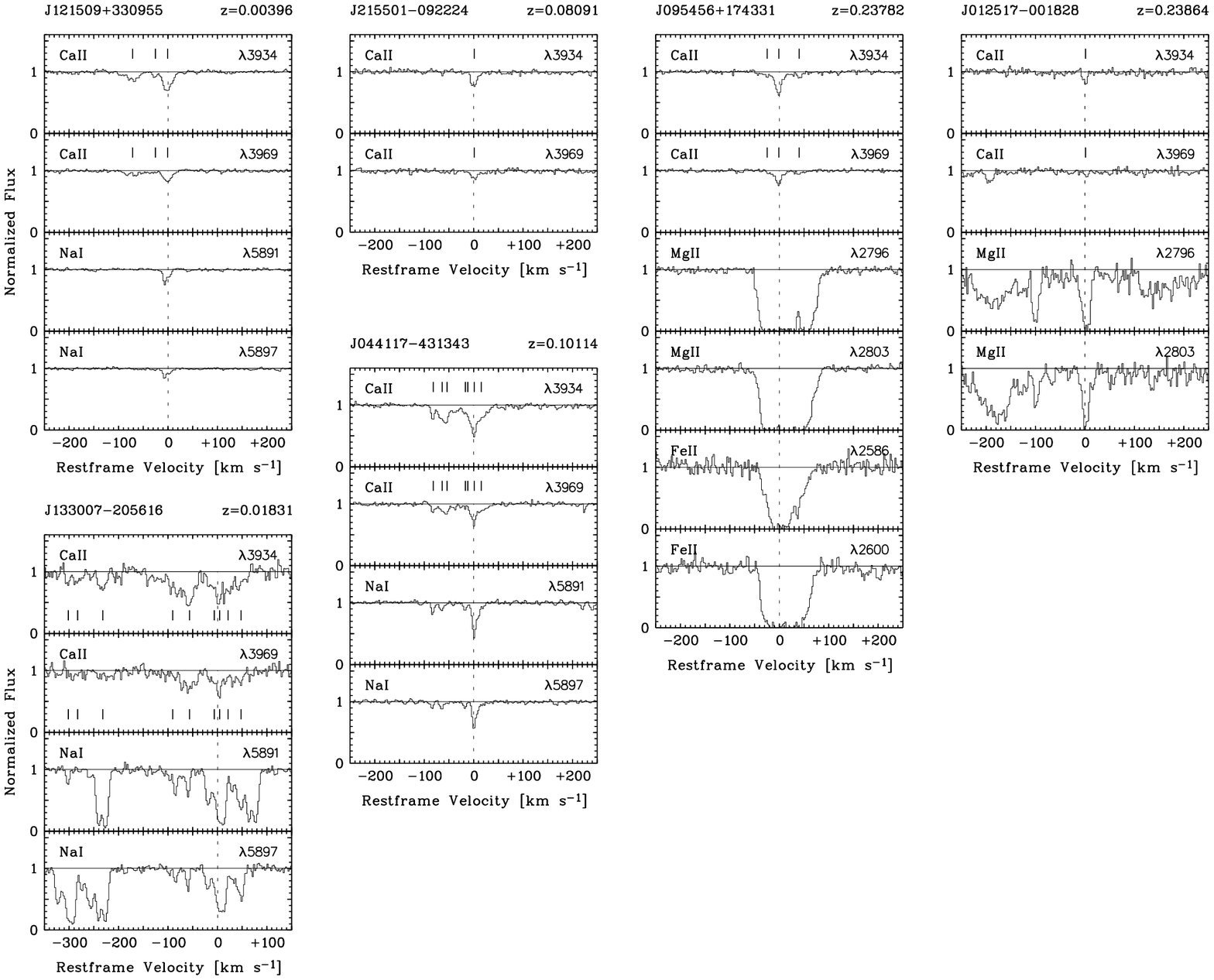}}
\caption[]{
Velocity profiles (absorber rest-frame) of Ca\,{\sc ii} and associated Na\,{\sc i},
Mg\,{\sc ii}, and Fe\,{\sc ii} absorption in various intervening Ca\,{\sc ii}
absorption systems. Quasar names and absorption redshifts are indicated at the
top of each panel.
}
\label{vel1}
\end{figure*}

%

\begin{figure*}[t!]
\centering
\resizebox{0.9\hsize}{!}{\includegraphics{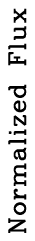}}
\caption[]{Same as for Fig.\,A.1.}
\label{vel2}
\end{figure*}

%

\begin{figure*}[t!]
\centering
\resizebox{0.9\hsize}{!}{\includegraphics{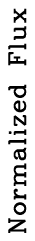}}
\caption[]{Same as for Fig.\,A.1.}
\label{vel3}
\end{figure*}

%

\begin{figure*}[t!]
\centering
\resizebox{0.9\hsize}{!}{\includegraphics{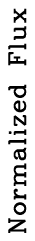}}
\caption[]{Same as for Fig.\,A.1.}
\label{vel4}
\end{figure*}

%

\begin{figure*}[t!]
\centering
\resizebox{0.9\hsize}{!}{\includegraphics{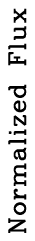}}
\caption[]{Same as for Fig.\,A.1.}
\label{vel5}
\end{figure*}

%

\begin{table*}[t!]
\caption[]{Complete list of QSOs in our data sample with mean S/N and Ca\,{\sc ii}
redshift path ($dz$), part 1}
\label{qsosample1}
\begin{scriptsize}
\begin{tabular}{rlrrrlrrrlrr}
\hline
No. & QSO & S/N$^{\rm a}$ & d$z$ &
No. & QSO & S/N$^{\rm a}$ & d$z$ &
No. & QSO & S/N$^{\rm a}$ & d$z$ \\
\hline
    1 &  J000344-232355  &  60 &  0.465 &            
    2 &  J000448-415728  &  35 &  0.315 &            
    3 &  J000815-095854  &  39 &  0.190 \\            
    4 &  J000852-290043  &  27 &  0.155 &            
    5 &  J000857-290126  &  18 &  0.080 &            
    6 &  J001210-012207  &  34 &  0.362 \\            
    7 &  J001306+000431  &  32 &  0.265 &            
    8 &  J001602-001225  &  47 &  0.500 &            
    9 &  J001643-575056  &  34 &  0.402 \\           
   10 &  J002110-242248  &  22 &  0.390 &            
   11 &  J002133+004301  &  26 &  0.390 &            
   12 &  J002151-012833  &  22 &  0.295 \\          
   13 &  J003023-181956  &  37 &  0.190 &            
   14 &  J004057-014632  &  23 &  0.500 &            
   15 &  J004131-493611  &  45 &  0.155 \\           
   16 &  J004201-403039  &  31 &  0.225 &           
   17 &  J004216-333754  &  19 &  0.325 &            
   18 &  J004428-243417  &  26 &  0.210 \\            
   19 &  J004508-291432  &  25 &  0.300 &            
   20 &  J004519-264050  &  22 &  0.053 &           
   21 &  J004612-293109  &  41 &  0.285 \\           
   22 &  J004812-255003  &  24 &  0.425 &            
   23 &  J004816-254745  &  26 &  0.375 &            
   24 &  J004848-260020  &  23 &  0.255 \\           
   25 &  J005024-252234  &  27 &  0.260 &            
   26 &  J005102-252846  &  20 &  0.320 &            
   27 &  J005109-255216  &  19 &  0.210 \\           
   28 &  J005127-280433  &  26 &  0.395 &            
   29 &  J005211-251857  &  31 &  0.265 &            
   30 &  J005419-254900  &  29 &  0.200 \\           
   31 &  J005758-264314  &  66 &  0.060 &            
   32 &  J005905+000651  &  22 &  0.500 &            
   33 &  J005925-411043  &  34 &  0.225 \\           
   34 &  J010104-285801  &  24 &  0.205 &            
   35 &  J010311+131617  &  91 &  0.220 &            
   36 &  J010516-184642  &  25 &  0.240 \\           
   37 &  J010604-254651  &  43 &  0.150 &            
   38 &  J010821+062327  &  91 &  0.187 &            
   39 &  J011143-350300  &  40 &  0.423 \\           
   40 &  J011504-302514  &  38 &  0.190 &            
   41 &  J011517-012704  &  22 &  0.500 &            
   42 &  J011818+025806  &  22 &  0.500 \\           
   43 &  J012031-270124  & 108 &  0.255 &            
   44 &  J012101+034412  &  25 &  0.500 &           
   45 &  J012227-042127  &  19 &  0.500 \\            
   46 &  J012303-005818  &  17 &  0.265 &            
   47 &  J012417-374423  &  76 &  0.465 &            
   48 &  J012517-001828  &  45 &  0.402 \\           
   49 &  J012528-000555  &  21 &  0.500 &            
   50 &  J012944-403346  &  62 &  0.367 &            
   51 &  J013105-213446  &  62 &  0.245 \\           
   52 &  J013233+011607  &  33 &  0.365 &            
   53 &  J013243-165448  &  39 &  0.500 &            
   54 &  J013301-400628  &  62 &  0.215 \\           
   55 &  J013405+005109  &  35 &  0.280 &            
   56 &  J013442-413611  &  30 &  0.130 &            
   57 &  J013754-270736  &  20 &  0.035 \\           
   58 &  J013825-000534  &  39 &  0.290 &            
   59 &  J013857-225447  &  64 &  0.500 &            
   60 &  J013901-082444  &  15 &  0.195 \\           
   61 &  J014125-092843  &  36 &  0.500 &            
   62 &  J014214+002324  &  52 &  0.110 &            
   63 &  J014333-391700  &  60 &  0.465 \\           
   64 &  J014631+133506  &  95 &  0.255 &            
   65 &  J014717+125808  &  22 &  0.380 &            
   66 &  J015318+000911  &  43 &  0.290 \\           
   67 &  J015327-431137  &  51 &  0.295 &            
   68 &  J015733-004824  &  23 &  0.365 &            
   69 &  J015741-104340  &  23 &  0.470 \\           
   70 &  J020157-113233  &  37 &  0.500 &            
   71 &  J020457-170119  &  47 &  0.500 &            
   72 &  J021741-370059  &  53 &  0.240 \\           
   73 &  J021857+081727  &  48 &  0.235 &            
   74 &  J022541+113425  &  13 &  0.230 &            
   75 &  J022620-285750  &  34 &  0.362 \\           
   76 &  J022928-364357  &  17 &  0.425 &            
   77 &  J023507-040205  &  21 &  0.500 &            
   78 &  J023838+163659  &  88 &  0.270 \\           
   79 &  J024008-230915  &  67 &  0.445 &            
   80 &  J024221+004912  &  21 &  0.500 &            
   81 &  J024312-055055  &  30 &  0.427 \\           
   82 &  J024658-123630  &  40 &  0.427 &            
   83 &  J025140-220026  &  56 &  0.145 &            
   84 &  J025240-553832  &  30 &  0.270 \\           
   85 &  J025607+011038  &  40 &  0.426 &            
   86 &  J025634-401300  &  36 &  0.415 &            
   87 &  J025927+074739  &  12 &  0.200 \\           
   88 &  J030000+004828  &  40 &  0.100 &            
   89 &  J030211-314030  &  20 &  0.315 &            
   90 &  J030249-321600  &  28 &  0.427 \\           
   91 &  J030844-295702  & 115 &  0.205 &            
   92 &  J031006-192124  &  39 &  0.410 &            
   93 &  J031009-192207  &  33 &  0.405 \\           
   94 &  J031257-563912  &  17 &  0.165 &            
   95 &  J033032-270438  &  33 &  0.415 &            
   96 &  J033106-382404  &  64 &  0.390 \\           
   97 &  J033108-252443  &  50 &  0.317 &            
   98 &  J033244-445557  &  49 &  0.260 &            
   99 &  J033413-400825  &  16 &  0.230 \\           
  100 &  J033626-201940  &  25 &  0.170 &            
  101 &  J033854-000521  &  19 &  0.230 &            
  102 &  J034943-381030  &  82 &  0.145 \\           
  103 &  J035128-142908  &  42 &  0.500 &            
  104 &  J035230-071102  &  32 &  0.500 &            
  105 &  J035320-231417  &  14 &  0.210 \\           
  106 &  J035405-272421  &  41 &  0.245 &            
  107 &  J040114-395132  &  23 &  0.290 &            
  108 &  J040353-360501  &  15 &  0.500 \\           
  109 &  J040718-441013  & 155 &  0.256 &            
  110 &  J041716-055345  &  31 &  0.500 &            
  111 &  J042214-384452  & 163 &  0.165 \\           
  112 &  J042353-261801  &  34 &  0.370 &            
  113 &  J042442-375620  &  16 &  0.500 &            
  114 &  J042644-520819  &  22 &  0.427 \\            
  115 &  J042707-130253  &  39 &  0.370 &            
  116 &  J042840-375619  &  17 &  0.230 &            
  117 &  J043037-485523  &  80 &  0.450 \\           
  118 &  J043403-435547  &  25 &  0.245 &            
  119 &  J044017-433308  &  20 &  0.260 &            
  120 &  J044026-163234  &  19 &  0.265 \\           
  121 &  J044117-431343  &  53 &  0.465 &            
  122 &  J044534-354704  &  35 &  0.150 &            
  123 &  J045214-164016  &  82 &  0.265 \\           
  124 &  J045313-130555  &  77 &  0.200 &            
  125 &  J045523-421617  &  26 &  0.335 &            
  126 &  J045608-215909  &  78 &  0.405 \\           
  127 &  J050112-015914  &  29 &  0.225 &            
  128 &  J050644-610941  &  18 &  0.500 &            
  129 &  J051410-332622  &  52 &  0.375 \\           
  130 &  J051707-441055  &  35 &  0.500 &            
  131 &  J051939-364611  &  20 &  0.425 &            
  132 &  J053007-250329  &  30 &  0.290 \\           
  133 &  J055158-211949  &  25 &  0.210 &            
  134 &  J055246-363727  &  39 &  0.500 &            
  135 &  J060008-504036  &  59 &  0.105 \\           
  136 &  J073918+013704  &  48 &  0.191 &            
  137 &  J081331+254503  & 252 &  0.285 &            
  138 &  J083052+241059  & 145 &  0.265 \\           
  139 &  J084205+183541  &  17 &  0.500 &            
  140 &  J084424+124546  &  29 &  0.105 &            
  141 &  J084650+052946  &  17 &  0.165 \\           
  142 &  J090910+012135  &  21 &  0.450 &            
  143 &  J091127+055054  & 102 &  0.280 &            
  144 &  J091127+055052A &  95 &  0.280 \\           
  145 &  J091613+070224  &  64 &  0.285 &            
  146 &  J092129-261843  &  59 &  0.405 &            
  147 &  J092507+144425  &  13 &  0.180 \\           
  148 &  J092913-021446  &  50 &  0.460 &            
  149 &  J093509-333237  &  24 &  0.035 &            
  150 &  J093518+020415  &  24 &  0.500 \\           
  151 &  J094253-110426  & 160 &  0.200 &            
  152 &  J094309+103400  &  23 &  0.322 &            
  153 &  J095352+080103  &  30 &  0.500 \\           
  154 &  J095456+174331  &  37 &  0.465 &            
  155 &  J100523+115712  &  20 &  0.265 &            
  156 &  J100731-333305  &  13 &  0.210 \\           
  157 &  J100930-002619  &  49 &  0.400 &            
  158 &  J101155+294141  &  63 &  0.325 &            
  159 &  J102837-010027  &  31 &  0.397 \\           
  160 &  J103909-231326  &  85 &  0.175 &            
  161 &  J104032-272749  &  53 &  0.385 &            
  162 &  J104033-272308  &  25 &  0.194 \\           
  163 &  J104040-272437  &  32 &  0.365 &            
  164 &  J104244+120331  &  15 &  0.285 &            
  165 &  J104540-101812  &  27 &  0.427 \\           
  166 &  J104642+053107  &  50 &  0.160 &            
  167 &  J104656+054150  &  64 &  0.245 &            
  168 &  J104733+052454  &  74 &  0.195 \\           
  169 &  J104800+052209  &  38 &  0.195 &            
  170 &  J105440-002048  &  49 &  0.402 &            
  171 &  J105800-302455  &  29 &  0.205 \\           
  172 &  J105817+195150  &  19 &  0.245 &            
  173 &  J110325-264515  & 212 &  0.465 &            
  174 &  J110729+004811  &  42 &  0.387 \\           
  175 &  J111350-153333  &  61 &  0.095 &            
  176 &  J111358+144226  &  22 &  0.500 &            
  177 &  J112229+180526  &  18 &  0.500 \\           
  178 &  J112442-170517  & 245 &  0.450 &            
  179 &  J112910-231628  &  19 &  0.410 &            
  180 &  J112932+020422  &  35 &  0.465 \\           
  181 &  J113007-144927  &  56 &  0.437 &            
  182 &  J114254+265457  & 163 &  0.137 &            
  183 &  J114608-244732  &  32 &  0.450 \\           
  184 &  J115411+063426  &  48 &  0.305 &            
  185 &  J115538+053050  &  41 &  0.075 &            
  186 &  J115944+011206  &  46 &  0.425 \\            
  187 &  J120044-185944  &  54 &  0.459 &            
  188 &  J120342+102831  &  34 &  0.457 &            
  189 &  J120550+020131  &  30 &  0.345 \\           
  190 &  J121140+103002  &  54 &  0.427 &            
  191 &  J121509+330955  & 125 &  0.427 &            
  192 &  J122310-181642  &  34 &  0.420 \\           
  193 &  J122607+173650  &  50 &  0.230 &            
  194 &  J123055-113909  &  47 &  0.040 &            
  195 &  J123200-022404  &  73 &  0.407 \\           
  196 &  J123313-102518  &  16 &  0.280 &            
  197 &  J123437+075843  &  98 &  0.327 &            
  198 &  J124524-000938  &  34 &  0.415 \\           
  199 &  J124604-073046  &  21 &  0.100 &            
  200 &  J124924-023339  &  71 &  0.370 &            
  201 &  J125151-022333  &  22 &  0.427 \\           
  202 &  J125212-132449  &  35 &  0.400 &            
  203 &  J125438+114105  &  23 &  0.500 &            
  204 &  J125838-180002  &  19 &  0.425 \\           
  205 &  J130753+064213  &  18 &  0.465 &            
  206 &  J131938-004940  &  16 &  0.260 &            
  207 &  J132113-263544  &  16 &  0.290 \\           
  208 &  J132323-002155  &  41 &  0.425 &            
  209 &  J132654-050058  &  22 &  0.427 &            
  210 &  J133007-205616  &  31 &  0.427 \\           
  211 &  J133335+164903  &  95 &  0.402 &            
  212 &  J134258-135559  &  71 &  0.165 &            
  213 &  J134427-103541  &  47 &  0.465 \\           
  214 &  J135038-251216  &  62 &  0.340 &            
  215 &  J135256-441240  &  46 &  0.465 &            
  216 &  J135704+191907  &  37 &  0.500 \\           
  217 &  J140445-013021  &  23 &  0.150 &            
  218 &  J141217+091624  &  51 &  0.274 &            
  219 &  J141520-095558  &  15 &  0.290 \\           
  220 &  J141803+170324  &  18 &  0.500 &            
  221 &  J141908+062834  &  15 &  0.450 &            
  222 &  J142249-272756  &  32 &  0.500 \\           
\hline
\end{tabular}
\noindent
\\
$^{\rm a}$\, S/N per resolution element.
\end{scriptsize}
\end{table*}

\begin{table*}[t!]
\caption[]{Complete list of QSOs in our data sample with mean S/N and Ca\,{\sc ii}
redshift path ($dz$), part 2}
\label{qsosample2}
\begin{scriptsize}
\begin{tabular}{rlrrrlrrrlrr}
\hline
No. & QSO & S/N$^{\rm a}$ & d$z$ &
No. & QSO & S/N$^{\rm a}$ & d$z$ &
No. & QSO & S/N$^{\rm a}$ & d$z$ \\
\hline
  223 &  J142253-000149  &  23 &  0.400 &            
  224 &  J142330+115951  &  18 &  0.500 &            
  225 &  J142738-120349  &  18 &  0.500 \\           
  226 &  J142906+011705  &  28 &  0.423 &            
  227 &  J143040+014939  &  32 &  0.291 &            
  228 &  J143229-010616A &  33 &  0.465 \\           
  229 &  J143649-161341  &  22 &  0.135 &            
  230 &  J143912+111740  &  60 &  0.240 &            
  231 &  J144653+011356  &  84 &  0.430 \\           
  232 &  J145102-232930  &  75 &  0.465 &            
  233 &  J145641-061743  &  23 &  0.500 &            
  234 &  J151053-054307  &  18 &  0.500 \\           
  235 &  J151126+090703  &  30 &  0.235 &            
  236 &  J151352+085555  & 126 &  0.235 &            
  237 &  J155035+052710  &  13 &  0.300 \\           
  238 &  J161749+024643  &  22 &  0.500 &            
  239 &  J162439+234512  &  41 &  0.402 &            
  240 &  J163145+115602  &  19 &  0.240 \\           
  241 &  J174358-035004  &  18 &  0.225 &            
  242 &  J194025-690756  &  98 &  0.174 &            
  243 &  J204719-163905  &  28 &  0.430 \\           
  244 &  J210244-355307  &  27 &  0.135 &            
  245 &  J211810-301911  &  26 &  0.500 &            
  246 &  J211927-353740  &  47 &  0.350 \\           
  247 &  J212329-005052  & 104 &  0.290 &            
  248 &  J212912-153841  & 232 &  0.150 &            
  249 &  J213302-464026  &  20 &  0.210 \\           
  250 &  J213314-464030  &  25 &  0.425 &            
  251 &  J213605-430818  &  35 &  0.265 &            
  252 &  J213638+004154  &  23 &  0.500 \\           
  253 &  J214222-441929  &  14 &  0.130 &            
  254 &  J214225-442018  &  37 &  0.150 &            
  255 &  J214622-152543  &  15 &  0.240 \\           
  256 &  J214805+065738  &  50 &  0.465 &            
  257 &  J215155-302753  &  43 &  0.355 &            
  258 &  J215324+053618  &  19 &  0.500 \\           
  259 &  J215501-092224  &  75 &  0.137 &            
  260 &  J215806-150109  &  17 &  0.500 &            
  261 &  J220734-403655  & 100 &  0.173 \\           
  262 &  J220743-534633  &  21 &  0.500 &            
  263 &  J220852-194359  & 141 &  0.400 &           
  264 &  J221511-004549  &  47 &  0.465 \\            
  265 &  J221531-174408  &  77 &  0.400 &            
  266 &  J221653-445156  &  45 &  0.414 &            
  267 &  J221852-033537  &  25 &  0.500 \\           
  268 &  J222006-280323  &  58 &  0.438 &            
  269 &  J222540-392436  &  25 &  0.425 &            
  270 &  J222547-045701  &  18 &  0.500 \\           
  271 &  J222756-224302  &  58 &  0.465 &            
  272 &  J222826-400957  &  20 &  0.390 &            
  273 &  J222940-083254  &  31 &  0.500 \\           
  274 &  J223951-294836  &  19 &  0.400 &            
  275 &  J224618-120651  &  29 &  0.500 &            
  276 &  J224708-601545  & 115 &  0.263 \\           
  277 &  J224752-123719  &  33 &  0.427 &            
  278 &  J225153-314619  &  44 &  0.060 &            
  279 &  J225154-314520  &  26 &  0.060 \\           
  280 &  J225310-365815  &  43 &  0.055 &            
  281 &  J225357+160853  &  73 &  0.450 &            
  282 &  J225805-275821  &  17 &  0.500 \\           
  283 &  J230001-341319  &  37 &  0.365 &            
  284 &  J231359-370446  &  40 &  0.262 &            
  285 &  J231646-404120  &  51 &  0.205 \\           
  286 &  J232046-294406  &  21 &  0.240 &            
  287 &  J232059-295521  &  14 &  0.150 &            
  288 &  J232115+142131  &  20 &  0.390 \\           
  289 &  J232121-294350  &  18 &  0.190 &            
  290 &  J232820+002238  &  52 &  0.415 &            
  291 &  J232917-473019  &  22 &  0.500 \\           
  292 &  J233544+150118  &  56 &  0.290 &            
  293 &  J233756-175220  &  22 &  0.500 &            
  294 &  J234256-032226  &  28 &  0.500 \\           
  295 &  J234628+124858  &  48 &  0.225 &            
  296 &  J234646+124526  &  60 &  0.205 &            
  297 &  J234646+124527  &  55 &  0.240 \\           
  298 &  J234819+005721A &  42 &  0.420 &            
  299 &  J234819+005717B &  22 &  0.365 &            
  300 &  J235034-432559  &  69 &  0.240 \\           
  301 &  J235057-005209  &  56 &  0.200 &            
  302 &  J235534-395355  &  36 &  0.425 &            
  303 &  J235731-112539  &  20 &  0.300 \\           
  304 &  J235953-124147  &  70 &  0.280 & 
      &                  &     &        &
      &                  &     &        \\           
\hline
\end{tabular}
\noindent
\\
$^{\rm a}$\, S/N per resolution element.
\end{scriptsize}
\end{table*}

\end{document}